\documentclass[12pt]{iopart}

\usepackage[dvipsnames]{xcolor}

\usepackage{graphicx}
\usepackage{tikz}

\usepackage{url}

\usepackage[normalem]{ulem} %

\usepackage{iopams} 
\newcommand{\mod}{\,\mathrm{mod}\,}

\newcommand{\rmeII}{\rho^{2\tau,\mathrm{MO}}_\eta}
\newcommand{\rmeIV}{\rho^{4\tau,\mathrm{MO}}_\eta}
\newcommand{\rmeVI}{\rho^{6\tau,\mathrm{MO}}_\eta}
\newcommand{\rmeVIII}{\rho^{8\tau,\mathrm{MO}}_\eta}
\newcommand{\rmpII}{\rho^{2\tau,\mathrm{MO}}_\eta}
\newcommand{\rmpIV}{\rho^{4\tau,\mathrm{MO}}_\eta}
\newcommand{\rmpVI}{\rho^{6\tau,\mathrm{MO}}_\eta}
\newcommand{\rmpVIII}{\rho^{8\tau,\mathrm{MO}}_\eta}
\newcommand{\rseII}{\rho^{2\tau,\mathrm{MO}}_\eta}
\newcommand{\rseIV}{\rho^{4\tau,\mathrm{MO}}_\eta}
\newcommand{\rseVI}{\rho^{6\tau,\mathrm{MO}}_\eta}
\newcommand{\rseVIII}{\rho^{8\tau,\mathrm{MO}}_\eta}
\newcommand{\rspII}{\rho^{2\tau,\mathrm{MO}}_\eta}
\newcommand{\rspIV}{\rho^{4\tau,\mathrm{MO}}_\eta}
\newcommand{\rspVI}{\rho^{6\tau,\mathrm{MO}}_\eta}
\newcommand{\rspVIII}{\rho^{8\tau,\mathrm{MO}}_\eta}
\newcommand{\sme}{\sigma^\mathrm{2MO}_\eta}
\newcommand{\smp}{\sigma^\mathrm{2MO}_\varphi}
\newcommand{\sse}{\sigma^\mathrm{2SC}_\eta}
\newcommand{\ssp}{\sigma^\mathrm{2SC}_\varphi}
\newcommand{\sdws}{\sigma^\mathrm{2DWS}}

\begin{document}

\title{Tilt-To-Length Coupling in LISA --- Uncertainty and Biases}

\author{M-S~Hartig$^{1,2,3}$, 
	J~Marmor$^{4,5}$, 
	D~George$^3$\footnote{Currently at the Wyant College of Optical Sciences at  The University of Arizona.}, 
	S~Paczkowski$^{2,1}$, 
	J~Sanjuan$^{4,5}$}

\address{$^1$ Leibniz Universit\"at Hannover, 30167 Hannover, Germany}
\address{$^2$ Max Planck Institute for Gravitational Physics (Albert-Einstein-Institut), 30167 Hannover, Germany}
\address{$^3$ Department of Physics, 2001 Museum Road, University of Florida, Gainesville, Florida 32611, USA}
\address{$^4$ Wyant College of Optical Sciences, The University of Arizona, 1630 E. University Blvd, Tucson, AZ 85719, USA}
\address{$^5$ Texas A\&M University, 701 H.R. Bright Bldg., College Station, TX 77843-3141, USA}

\ead{marie-sophie.hartig@aei.mpg.de}

\begin{abstract}
The coupling of the angular jitter of the spacecraft and their sub-assemblies with the optical bench and the telescope into the interferometric length readout will be a major noise source in the LISA mission. We refer to this noise as tilt-to-length (TTL) coupling.
It will be reduced directly by realignments, and the residual noise will then be subtracted in post-processing. 
The success of these mitigation strategies depends on an accurate computation of the TTL coupling coefficients. 
We present here a thorough analysis of the accuracy of the coefficient estimation under different jitter characteristics, angular readout noise levels, and gravitational wave sources. 
We analyze in which cases the estimates degrade using two estimators, the common least squares estimator and the instrumental variables estimator.
Our investigations show that angular readout noise leads to a systematic bias of the least squares estimator, depending on the TTL coupling coefficients, jitter and readout noise level, while the instrumental variable estimator converges to an unbiased result as the data set length increases. 
We present an equation that predicts the estimation bias of the least squares method due to angular readout noise. 
\end{abstract}

\section{Introduction}

The laser interferometer space observatory (LISA) will be the first detector to measure gravitational waves (GWs) in the milli-Hertz regime \cite{Danzmann2011,Redbook2024}.
It will consist of three spacecraft (SC) on heliocentric orbits. They are arranged in a triangular constellation and will exchange laser beams over $2.5\cdot10^9$\,m. 
The GW signals will be encoded in the distance changes between the SC and LISA will measure them using laser interferometry. Specifically, it will detect the minuscule distance changes between six test masses in free fall, two of which are located inside each SC. 
In early 2024, LISA was officially adopted by the European Space Agency (ESA) and thus it has transitioned from the study to the implementation phase~\cite{Redbook2024,LISAadoption}. 
At the time of writing, LISA is in the mission phase B2, which involves a detailed definition of the design and parallel performance studies.

Our manuscript addresses the characterization and suppression of one of the major noise sources in LISA, tilt-to-length (TTL) coupling~\cite{Hartig22,Hartig23}. 
TTL noise originates from the angular jitter of the SC and their moving optical sub-assemblies (MOSAs) including the optical benches and the telescopes, which  translates into interferometric length measurements.
The magnitude of the coupling depends on optical alignments, the beam's wavefront properties and the jitter amplitudes.
The intra-satellite TTL coupling can be explained by mechanisms similar to those observed during the LISA Pathfinder (LPF) mission~\cite{Hartig2023_LPF,Armano2023_TTL} and its minimization using imaging systems has been studied \cite{Chwalla2016,Troebs2018,Chwalla2020}.
A major difference of the TTL coupling between the LPF and the LISA mission are the TTL effects due to wavefront errors in the far-field. A more detailed characterization of this TTL coupling type can be found in \cite{Weaver2022,Sasso2019,Sasso2018,Xiao2023,Wang2024,Wang2020,Lin2023}.
In LISA, a direct TTL noise reduction by realignment is planned after launch, while residual TTL noise will be suppressed in post-processing~\cite{Wanner2024,Paczkowski2022,George2022,Houba2022a,Houba2022b,Wegener2024}.
For the TTL coupling estimation and suppression in post-processing, TTL coupling coefficients that scale the coupling of the angular jitter to the length measurement will be computed. The fitted TTL noise model will then be subtracted.
We analyze the accuracy of TTL coupling coefficient estimation as a function of jitter levels and shapes, as well as angular readout noise levels.
Our investigations focus on the following questions:
How does the accuracy of the coefficient estimation depend on the jitter and readout noise characteristics?
How can we minimize the angular readout noise bias in the coefficient estimates? 
How do GW signals affect the estimation accuracy?
The questions are of particular interest for TTL noise suppression by in-orbit realignment. This mitigation strategy relies on a precise computation of the coupling coefficients.  
While underestimating the TTL coupling coefficients may in some cases result in a lower noise residual since less angular readout noise would be added to the data \cite{Paczkowski2022,Armano2023_TTL}, such a bias must be minimized when using the coefficients for a realignment strategy.

The paper is organized as follows.
In \sref{sec:TTL_coupling}, we give a brief introduction of TTL coupling, and define it for the LISA mission in~\sref{sec:TTL_LISA}.
In \sref{sec:TTL_estimators}, we present the TTL coupling estimators used in this work. 
The performance of these estimators is analyzed first for different jitter and readout noise levels in \sref{sec:TTL_uncertainty}, and then for different GW sources in the measurement band in \sref{sec:TTL_GW}.
In \sref{sec:summary}, we summarize our results.

\section{Tilt-to-length coupling}
\label{sec:TTL_coupling}

In precision laser interferometry, tilt-to-length (TTL) coupling describes the inadvertent coupling of motion, that is orthogonal to the optical beam axis, into the phase readout \cite{Hartig22,Hartig23}.
We distinguish between angular and translational (`lateral') jitter coupling. 
We refer to lateral jitter coupling as TTL noise since lateral motion couples into the longitudinal measurement only in the case of angular misalignment. 
Lateral jitter yielded significant TTL noise during the LPF mission \cite{Armano2016,Hartig2023_LPF,Armano2023_TTL}.
However, it can be assumed negligible in LISA as explained in~\cite{Wanner2024}.
Furthermore, the TTL coupling comprises geometric, i.e.\ propagation time changes of the beam ray \cite{Hartig22}, and non-geometric, i.e.\ beam or detector properties related coupling \cite{Hartig23}, contributions.
In the linear case, TTL noise is given by an equation like
\begin{eqnarray}
  N^\mathrm{TTL} = \sum_i C^\mathrm{TTL}_i \cdot \alpha_i \,,
\label{eq:TTL_simple}
\end{eqnarray}
where $C^\mathrm{TTL}_i$ [mm/rad (or unitless)] are the coupling coefficients depending on the beam and optical setup properties. It scales the coupling of angular (or lateral) jitter $\alpha_i$ into the longitudinal readout.

In practice, the TTL noise $N^\mathrm{TTL}$ cannot be estimated without some uncertainty. 
The accuracy of the estimated noise $\hat{N}^\mathrm{TTL}$ is limited by the error $\delta C^\mathrm{TTL}$ of the computed coupling coefficient $\widehat{C}$ and the readout noise $\delta \alpha$ of the jittering variable. We have
\begin{eqnarray}
  \widehat{N}^\mathrm{TTL} &= \sum_i \widehat{C}^\mathrm{TTL}_i \cdot \widehat{\alpha}_i \nonumber \\
  &= \sum_i \left(C^\mathrm{TTL}_i+\delta C^\mathrm{TTL}_i\right) \cdot \left(\alpha_i+\delta\alpha_i\right) \nonumber \\
  &\approx N^\mathrm{TTL} + \sum_i \left(\delta C^\mathrm{TTL}_i\,\alpha_i + C^\mathrm{TTL}_i\,\delta\alpha_i \right) \,.
\label{eq:TTL_simple_error}
\end{eqnarray} 
where \ $\widehat{ }$ \ indicates measured or estimated. 
Note that the last equation in \eref{eq:TTL_simple_error} neglects the second-order error terms, which are significantly smaller than the shown linear contributions.
For simplicity, we will omit the superscript of $C^\mathrm{TTL}_i$ in the following.

\section{Tilt-to-length noise in LISA}
\label{sec:TTL_LISA}

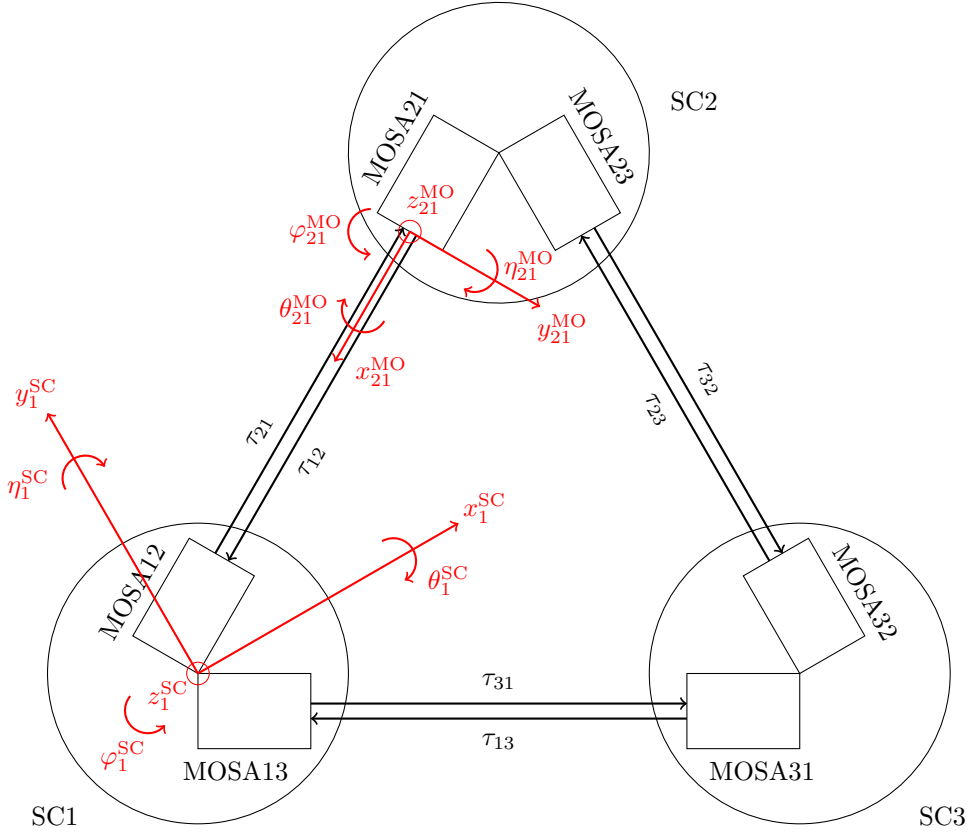
\begin{figure}
\begin{indented}
\item[]\begin{tikzpicture}
\draw (0,0) circle (2);
\draw (-1.9,-1.9) node {SC1};
\draw[rotate around={60:(0,0)}] (8,0) circle (2);
\draw[rotate around={60:(0,0)}] (9.9,-1.9) node {SC2};
\draw (8,0) circle (2);
\draw (9.9,-1.9) node {SC3};
\draw[rotate around={60:(0,0)}] (0,0) rectangle (1.5,1);
\draw[rotate around={60:(0,0)}] (0.5,1.3) node[rotate around={60:(0,0)}] {MOSA12}; 
\draw (0,-1) rectangle (1.5,0);
\draw (0.5,-1.3) node {MOSA13};
\draw[rotate around={60:(0,0)}] (6.5,0) rectangle (8,1);
\draw[rotate around={60:(0,0)}] (7.5,1.3) node[rotate around={60:(0,0)}] {MOSA21}; 
\draw[rotate around={-60:(8,0)}] (0,0) rectangle (1.5,1);
\draw[rotate around={-60:(8,0)}] (0.5,1.3) node[rotate around={-60:(0,0)}] {MOSA23}; 
\draw[rotate around={30:(8,0)}] (8,0) rectangle (9,1.5);
\draw[rotate around={-60:(8,0)}] (7.5,1.3) node[rotate around={-60:(0,0)}] {MOSA32}; 
\draw (6.5,-1) rectangle (8,0);
\draw (7.5,-1.3) node {MOSA31};
\draw[thick, ->,rotate around={60:(0,0)}] (1.5,0.6) -- (6.5,0.6);
\draw[rotate around={60:(0,0)}] (3.2,0.9) node[rotate around={60:(0,0)}] {$\tau_{21}$};
\draw[thick, ->,rotate around={60:(0,0)}] (6.5,0.4) -- (1.5,0.4);
\draw[rotate around={60:(0,0)}] (3.2,0.1) node[rotate around={60:(0,0)}] {$\tau_{12}$};
\draw[thick, ->] (1.5,-0.4) -- (6.5,-0.4);
\draw (4,-0.1) node {$\tau_{31}$};
\draw[thick, ->] (6.5,-0.6) -- (1.5,-0.6);
\draw (4,-0.9) node {$\tau_{13}$};
\draw[thick, ->,rotate around={-60:(8,0)}] (6.5,0.4) -- (1.5,0.4);
\draw[rotate around={-60:(8,0)}] (4,0.1) node[rotate around={-60:(0,0)}] {$\tau_{23}$};
\draw[thick, ->,rotate around={-60:(8,0)}] (1.5,0.6) -- (6.5,0.6);
\draw[rotate around={-60:(8,0)}] (4,0.9) node[rotate around={-60:(0,0)}] {$\tau_{32}$};
\draw[red, thick, ->,rotate around={30:(0,0)}] (0,0) -- (4,0); 
\draw[red,rotate around={30:(0,0)}] (4.4,0) node {$x_{1}^{\rm SC}$};
\draw[red, thick, ->,rotate around={30:(0,0)}] (0,0) -- (0,4); 
\draw[red,rotate around={30:(0,0)}] (0,4.3) node {$y_{1}^{\rm SC}$};
\draw[red] (0,0) circle (0.15);
\draw[red] (-0.4,-0.3) node {$z_{1}^{\rm SC}$};
\draw[red, thick, ->,rotate around={30:(0,0)}] (3,0.3) arc (90:-90:0.3);
\draw[red,rotate around={30:(0,0)}] (3.5,-0.6) node {$\theta_{1}^{\rm SC}$};
\draw[red, thick, ->,rotate around={30:(0,0)}] (-0.3,3) arc (180:0:0.3);
\draw[red,rotate around={30:(0,0)}] (-0.7,3.3) node {$\eta_{1}^{\rm SC}$};
\draw[red, thick, ->] (-0.9,-0.3) arc (140:320:0.3);
\draw[red] (-1,-1.1) node {$\varphi_{1}^{\rm SC}$};
\draw[red, thick, ->,rotate around={60:(0,0)}] (6.5,0.5) -- (4.5,0.5);
\draw[red, rotate around={60:(0,0)}] (4.7,-0.1) node {$x_{21}^{\rm MO}$};
\draw[red, thick, ->,rotate around={60:(0,0)}] (6.5,0.5) -- (6.5,-1.5);
\draw[red, rotate around={60:(0,0)}] (6.4,-1.9) node {$y_{21}^{\rm MO}$};
\draw[red, rotate around={60:(0,0)}] (6.5,0.5) circle (0.15);
\draw[red] (3.1,6.3) node {$z_{21}^{\rm MO}$};
\draw[red, thick, ->,rotate around={60:(0,0)}] (5.3,0.2) arc (270:90:0.3);
\draw[red,rotate around={60:(0,0)}] (4.9,1.2) node {$\theta_{21}^{\rm MO}$};
\draw[red, thick, ->,rotate around={60:(0,0)}] (6.8,-0.5) arc (360:180:0.3);
\draw[red,rotate around={60:(0,0)}] (6.9,-1.1) node {$\eta_{21}^{\rm MO}$};
\draw[red, thick, ->,rotate around={60:(0,0)}] (6.5,1.1) arc (30:210:0.3);
\draw[red,rotate around={60:(0,0)}] (5.9,1.6) node {$\varphi_{21}^{\rm MO}$};
\end{tikzpicture}
\end{indented}
\caption{Naming conventions for the LISA SC and MOSAs. Each SC and MOSA has its own coordinate system. We show them exemplary for SC1 and MOSA21. The $\tau_{ij},\ i,j\in\{1,2,3\}$ are the light propagation times between the SC.}
\label{fig:LISA_notation}
\end{figure}

In LISA, TTL coupling predominately originates from the angular jitter of the SC and the MOSAs, see \fref{fig:LISA_notation}. 
Overall, TTL coupling is expected to be one of the major noise sources in LISA. 
In this section, we introduce the contributing TTL coupling terms and the TTL noise characterization in post-processing. 

\subsection{Tilt-to-length coupling for a single link}

The TTL coupling in LISA's long-arm or inter-satellite interferometer is driven by alignment imperfections and wavefront errors \cite{Wanner2024}. 
These couple into the length measurements via the jitter of the local and the far SC. 
The TTL noise measured by the interferometer on SC $i$ receiving light from SC $j$ is \cite{Wanner2024,Paczkowski2022}
\begin{eqnarray}
	N^\mathrm{TTL}_{ij} = 
	C_{ij\varphi \rm RX} \, \varphi_{ij} +
	C_{ij\eta \rm RX} \, \eta_{ij} +
	C_{ji\varphi \rm TX} \, D_{ij} \varphi_{ji} +
	C_{ji\eta \rm TX} \, D_{ij} \eta_{ji} \,.
\label{eq:TTL_singlelink}
\end{eqnarray}
Here, $\varphi_{ij}$ and $\eta_{ij}$ are the yaw and pitch jitter of the `receiving' SC~$i$ and its MOSA pointing towards SC $j$.
The jitter of the `transmitting' SC~$j$ and its MOSA pointing towards SC $i$ have to be delayed by the light traveling time $\tau_{ij}$. $D_{ij}$ are the corresponding delay operators. 
In total, there are six links between the SC, $ij\in\{12,13,23,21,31,32\}$. This yields in total 24 TTL coupling terms.
The notation of the angles is exemplary shown in \fref{fig:LISA_notation} for SC~1 and MOSA~21.

\subsection{Time delay interferometry combinations}

It is not possible to study TTL coupling for a single link due to the preeminent laser frequency noise, which is the primary noise in LISA. 
The laser frequency noise will be suppressed by time-delay interferometry (TDI) \cite{Tinto2021,Tinto2023}. By the combination of delayed length measurements, a virtual equal arm interferometer is constructed. We will need 2nd generation TDI combinations to account for changing LISA arm lengths.
While there are multiple possible TDI combinations \cite{Tinto2004,Shaddock2003,Muratore2020}, we will focus on the often used 2nd generation Michelson TDI combination \cite{Tinto2021}.
The TTL noise terms have to be traced through this TDI combination as well. 
For the TDI\,X combination, we find
\begin{eqnarray}
	N^\mathrm{TTL}_\mathrm{TDIX} &=&  
	\left(1 -D_{12}D_{21} -D_{12}D_{21}D_{13}D_{31} +D_{13}D_{31}D_{12}D_{21}D_{12}D_{21}\right) \nonumber\\
	&&\cdot \left( C_{13\eta{\rm RX}}\,\eta_{13} + C_{31\eta{\rm TX}}\,D_{13}\eta_{31} + C_{31\eta{\rm RX}}\,D_{13}\eta_{31} + C_{13\eta{\rm TX}}\,D_{131}\eta_{13} \right. \nonumber\\
	&&\ \ \left. +C_{13\varphi{\rm RX}}\,\varphi_{13} + C_{31\varphi{\rm TX}}\,D_{13}\varphi_{31} + C_{31\varphi{\rm RX}}\,D_{13}\varphi_{31} + C_{13\varphi{\rm TX}}\,D_{131}\varphi_{13} \right) \nonumber\\
	&-& \left(1 -D_{13}D_{31} -D_{13}D_{31}D_{12}D_{21} +D_{12}D_{21}D_{13}D_{31}D_{13}D_{31}\right) \nonumber\\
	&&\cdot \left( C_{12\eta{\rm RX}}\,\eta_{12} + C_{21\eta{\rm TX}}\,D_{12}\eta_{21} + C_{21\eta{\rm RX}}\,D_{12}\eta_{21} + C_{12\eta{\rm TX}}\,D_{121}\eta_{12}  \right. \nonumber\\
	&&\ \ \left. +C_{12\varphi{\rm RX}}\,\varphi_{12} + C_{21\varphi{\rm TX}}\,D_{12}\varphi_{21} + C_{21\varphi{\rm RX}}\,D_{12}\varphi_{21} + C_{12\varphi{\rm TX}}\,D_{121}\varphi_{12} \right) \,.
\label{eq:TTL_TDI}
\end{eqnarray}
TDI\,Y and TDI\,Z combinations can be found by cyclical permutations of the indices. 
A detailed modeling of TTL coupling in the single links and the TDI Michelson observables is presented in \cite{Wanner2024}.

\subsection{Tilt-to-length coupling characterization}

TTL coupling noise will be reduced by design and realignment prior to launch.
The remaining in-orbit noise will be fitted and subtracted in post-processing.
LPF data analyses \cite{Armano2023_TTL,Wanner2017} and studies with simulated LISA data \cite{Paczkowski2022,George2022} have shown that TTL coupling can be estimated and subtracted from the mission data on a daily basis.
In addition, maneuver designs for a better coupling coefficient estimation have been investigated \cite{Houba2022a,Houba2022b,Wegener2024}. 

As motivated in \sref{sec:TTL_coupling}, the accuracy of the estimated TTL coupling noise is limited by the coupling coefficient estimation error and the angular readout noise. 
We will show in the following that the TTL coefficient error depends on the instrument noise in the inter-satellite interferometer readout, the angular readout noise, and the GW signals.
Some of these dependencies have been previously investigated in \cite{Paczkowski2022,George2022}.
By subtracting the estimated TTL coupling $\widehat{N}^\mathrm{TTL}$ from the length data, we add in angular readout noise $\delta \alpha$. The readout noise will be scaled by the coupling coefficients, see equation~\eref{eq:TTL_simple_error}.
This effect has also been observed in LPF \cite{Armano2023_TTL,Wanner2017} and requires for an upper bound on the coupling coefficients.
The TTL noise subtraction scheme presented in \cite{Paczkowski2022} accounts for the addition of angular readout noise. There, a Markov Chain Monte Carlo (MCMC) method is used to minimize the residual after subtraction.
Unfortunately, this method yields less accurate TTL coupling coefficient estimates as the angular readout noise becomes large.
Precise coupling coefficients are needed to characterize the TTL coupling noise, e.g.\ when in-flight realignment for TTL coupling suppression becomes necessary.

A key question in this paper is how well the coefficients can be estimated for different levels of angular jitter and readout noise, and in the presence of as GW signals.
Since a linear TTL coupling model is considered being sufficient for TTL coupling suppression below the other instrument noises \cite{Wanner2024,Paczkowski2022}, we restrict our analysis to linear models. 
I.e., this manuscript does not cover higher-order TTL coupling contributions or coupling coefficient drifts.
For a characterization of second-order TTL coupling effects in the far-field, see \cite{Weaver2022} for the LISA mission and \cite{Xiao2023,Wang2024} for other space-based GW detector concepts.
TTL coupling coefficient drifts were previously investigated in \cite{Paczkowski2022}.

\section{Tilt-to-length noise estimators}
\label{sec:TTL_estimators}

For the estimation of the TTL coupling coefficients, we use the interferometric measurements of the length changes between two test masses and the angular readouts of the inter-satellite interferometer.
The length readout for a single link is given by
\begin{eqnarray}
  S^\mathrm{IFO}_{ij} = N^\mathrm{TTL}_{ij} + N^\mathrm{Instr}_{ij} + H^\mathrm{GW}_{ij} \,,
\label{eq:signal_ISI}
\end{eqnarray}
with 
\begin{description}
  \item[$N^\mathrm{TTL}_{ij}$ :] TTL coupling noise as defined in equation~\eref{eq:TTL_singlelink}. 
  \item[$N^\mathrm{Instr}_{ij}$ :] all other instrument noise sources adding to the readout. A complete list of the noises considered in our analysis is given in \ref{app:ifo_noise}.
  \item[$H^\mathrm{GW}_{ij}$ :] the GW signals.
\end{description}
The combined angular SC and MOSA jitter will be measured utilizing differential wavefront sensing (DWS) \cite{Morrison1994,Wanner2012}. For the idealized case of an constant opening angle of 30°, which we would find for equal arm lengths, the DWS measurement can be described by the equations.
\begin{eqnarray}
  \varphi^\mathrm{DWS}_{ij} =& 
  \varphi^\mathrm{SC}_i + \varphi^\mathrm{MO}_{ij} + \delta\varphi_{ij} \,, 
  \label{eq:signal_DWS_phi} \\
  \eta^\mathrm{DWS}_{ij} =& 
  \frac{\sqrt{3}}{2}\eta^\mathrm{SC}_i + (-1)^k\ \frac{1}{2}\theta^\mathrm{SC}_i + \eta^\mathrm{MO}_{ij} + \delta\eta_{ij} 
  \label{eq:signal_DWS_eta} \\
  &\ \mathrm{with}\ k=\cases{1 & for \ $ij\in\{12,23,31\}$ \\ 0 & for \ $ij\in\{13,21,32\}$} \,. \nonumber
\end{eqnarray}
In fact, the opening angle in LISA can vary by up to $\pm$1° \cite{martens2021}. This would change the factors of $\eta^\mathrm{SC}_i$ and $\theta^\mathrm{SC}_i$ in equation~\eref{eq:signal_DWS_eta} by a maximum of 3\%, which will be ignored in the following analysis.
Note further that we cannot distinguish between SC and MOSA jitter. However, the coupling coefficients for both jitters can be considered to be similar \cite{Wanner2024}.
$\delta\alpha_{ij},\ \alpha\in\{\varphi,\eta\}$ is the DWS readout noise.
Both the longitudinal and angular readouts must be traced through TDI for the TTL coefficient computation.

For the estimation of the TTL coupling coefficients, we have to find a solution that minimizes the equation
\begin{eqnarray}
    R_\mathrm{TDI} &=& S_\mathrm{TDI}^\mathrm{IFO} - \widehat{N}_\mathrm{TDI}^\mathrm{TTL} \,,
\end{eqnarray}
where
\begin{eqnarray}
	S_\mathrm{TDI}^\mathrm{IFO} &=&  
	\left(1 -D_{12}D_{21} -D_{12}D_{21}D_{13}D_{31} +D_{13}D_{31}D_{12}D_{21}D_{12}D_{21}\right) \nonumber\\
    && \cdot \left( S^\mathrm{IFO}_{13} + D_{13}\,S^\mathrm{IFO}_{31} \right) \nonumber\\
	&-& \left(1 -D_{13}D_{31} -D_{13}D_{31}D_{12}D_{21} +D_{12}D_{21}D_{13}D_{31}D_{13}D_{31}\right) \nonumber\\
    && \cdot \left( S^\mathrm{IFO}_{12} + D_{12}\,S^\mathrm{IFO}_{21} \right)
\label{eq:ISI_TDI}
\end{eqnarray}
is the interferometric single link length readout (equation~\eref{eq:signal_ISI}) propagated through TDI\,X. The TDI\,Y and TDI\,Z variables can be found by cyclical permutations of the indices.
Furthermore, we have
\begin{description}
  \item[$\widehat{N}_\mathrm{TDI}^\mathrm{TTL}$ :] TTL coupling noise model after TDI, which refers for TDI\,X to equation~\eref{eq:TTL_TDI} with the angular jitter being replaced by the DWS measurements from equation~\eref{eq:signal_DWS_phi} and \eref{eq:signal_DWS_eta}.
  \item[$R_\mathrm{TDI}$ :] residual after subtraction of the estimated TTL coupling from the length measurement after TDI.
\end{description}
Given the length and angular measurements described above, there are several possible estimators for the TTL coupling coefficient estimation~\cite{Pintelon2001}. 
A common parameter estimation method is the linear least squares (LS) estimator. This estimator is fast and computationally less expensive than other methods. Under certain conditions, it coincides with the maximum likelihood estimates.
However, estimation results become biased if the independent variable, i.e.\ the angular jitter measurement, is noisy. 
Therefore, we also investigate the instrumental variable (IV) estimator. This estimator is unbiased under our assumptions and does not require more computational resources than the LS estimator. 
Since we are dealing with long data sets and 24 coupling coefficients, other methods quickly become very expensive due to the large matrices and vectors involved. 
In the following, we will introduce both estimators and their convergence properties.

\subsection{Least squares}
\label{sec:TTL_estimators_LS}

Let there be data sets of the inter-satellite length change and angular readouts for times $\mathbf{t}\in\mathbb{R}^N$.
Then, the LS algorithm computes the TTL coupling coefficients via
\begin{eqnarray}
  \widehat{\mathbf{C}}^\mathrm{LS}
  = \left(\sum_{n=1}^{3N} \left[\balpha_\mathrm{TDI}^\mathrm{DWS}(t_n)\right]^T\cdot\balpha_\mathrm{TDI}^\mathrm{DWS}(t_n)\right)^{-1}
  \cdot \left(\sum_{n=1}^{3N} \left[\balpha_\mathrm{TDI}^\mathrm{DWS}(t_n)\right]^T\cdot \mathbf{S}_\mathrm{TDI}^\mathrm{IFO}(t_n)\right)
  \label{eq:LS}
\end{eqnarray}
with 
\begin{description}
  \item[$\mathbf{S}_\mathrm{TDI}^\mathrm{IFO}(t_n)\in\mathbb{R}$ :] TDI\,X (if $n\leq N$), TDI\,Y (if $N<n\leq2N$), TDI\,Z (if $n>2N$) combination of the signals $S^\mathrm{IFO}_{ij}$ given by equation~\eref{eq:signal_ISI}. For TDI\,X, compare with equation~\eref{eq:ISI_TDI}.
  \item[$\balpha_\mathrm{TDI}^\mathrm{DWS}(t_n)\in\mathbb{R}^{1\times24}$ :] vector of the TDI\,X (if $n\leq N$), TDI\,Y (if $N<n\leq2N$), TDI\,Z (if $n>2N$) combinations of the 12 angular readouts $\alpha^\mathrm{DWS}_{ij}$, $\alpha\in\{\varphi,\eta\}$, given by equations~\eref{eq:signal_DWS_phi} and \eref{eq:signal_DWS_eta} and their 12 delays along the corresponding LISA arm $D_{ji}\,\alpha^\mathrm{DWS}_{ij}$. The computation of the TDI variables of the angular measurement is equivalent to the respective computation for the $S^\mathrm{IFO}_{ij}$ readouts (see equation~\eref{eq:ISI_TDI}).
  \item[$t_n$ :] time stamp. It is $t_n=T_s\cdot[(n \mod N)-1]$, where $T_s$ is the sampling period.
\end{description}

If the readout error of the angular measurements $\delta\balpha_\mathrm{TDI}$ is zero, the LS estimator converges to the true TTL coupling coefficient:
\begin{eqnarray}\hspace*{-\mathindent}
  \lim_{N\rightarrow\infty} \widehat{\mathbf{C}}^\mathrm{LS}
  &=& \lim_{N\rightarrow\infty} \left(\sum_{n=1}^{3N} \left[\balpha_\mathrm{TDI}(t_n)\right]^T\cdot\balpha_\mathrm{TDI}(t_n)\right)^{-1}
  \cdot \left(\sum_{n=1}^{3N} \left[\balpha_\mathrm{TDI}(t_n)\right]^T\cdot \mathbf{S}_\mathrm{TDI}^\mathrm{IFO}(t_n)\right) \nonumber \\
  &=& \lim_{N\rightarrow\infty} \left(\sum_{n=1}^{3N} \left[\balpha_\mathrm{TDI}(t_n)\right]^T\cdot\balpha_\mathrm{TDI}(t_n)\right)^{-1} \nonumber \\
  &&\quad \cdot \left(\sum_{n=1}^{3N} \left[\balpha_\mathrm{TDI}(t_n)\right]^T\cdot \left[\mathbf{N}_\mathrm{TDI}^\mathrm{TTL}(t_n)+\mathbf{N}_\mathrm{TDI}^\mathrm{Instr}(t_n)+\mathbf{H}_\mathrm{TDI}^\mathrm{GW}(t_n)\right]\right) \nonumber\\
  &=& \lim_{N\rightarrow\infty} \left(\sum_{n=1}^{3N} \left[\balpha_\mathrm{TDI}(t_n)\right]^T\cdot\balpha_\mathrm{TDI}(t_n)\right)^{-1}
  \cdot \left(\sum_{n=1}^{3N} \left[\balpha_\mathrm{TDI}(t_n)\right]^T\cdot \left[\textbf{C}\cdot\balpha_\mathrm{TDI}(t_n)\right]\right) \nonumber\\
  &=&\mathbf{C} \,. 
\end{eqnarray}
Here we assume that the angular jitter and the other instrument noise as well as the GW signals are uncorrelated.

In realistic scenarios, the angular readout error will not be zero. The data will be affected by shot noise, phasemeter noise and other noise sources \cite{Armano2022_OMS}. Considering angular readout noise in our analysis and calculating the error of the LS estimate (i.e.\ computing $\lim_{N\rightarrow\infty} \widehat{\mathbf{C}}^\mathrm{LS}-\mathbf{C}$), we find for white jitter and white readout noise
\begin{eqnarray}\hspace*{-\mathindent}
  \lim_{N\rightarrow\infty} \delta \widehat{C}^\mathrm{LS}_{ij \eta \rm RX} 
  &=& \frac{\sdws/2}{2\,\sme+\sse+2\,\sdws} \left( C_{ij \eta \rm RX} +C_{ik \eta \rm RX} -C_{ij \eta \rm TX} -C_{ik \eta \rm TX} \right) \nonumber\\
  &+& \frac{\sdws/2}{2\,\sme+3\,\sse+2\,\sdws} \left( C_{ij \eta \rm RX} -C_{ik \eta \rm RX} -C_{ij \eta \rm TX} +C_{ik \eta \rm TX} \right) \nonumber\\
  &+& \frac{13\,\sdws/2}{26\,\sme+25\,\sse+26\,\sdws} \left( C_{ij \eta \rm RX} +C_{ik \eta \rm RX} +C_{ij \eta TX} +C_{ik \eta \rm TX} \right) \nonumber\\
  &+& \frac{13\,\sdws/2}{26\,\sme+27\,\sse+26\,\sdws} \left( C_{ij \eta RX} -C_{ik \eta RX} +C_{ij \eta TX} -C_{ik \eta TX} \right) \nonumber \\ 
  \,,
  \label{eq:dC_etaRX}\\ \hspace*{-\mathindent}
  \lim_{N\rightarrow\infty} \delta \widehat{C}^\mathrm{LS}_{ij \eta \rm TX} 
  &=& \frac{\sdws/2}{2\,\sme+\sse+2\,\sdws} \left( -C_{ij \eta \rm RX} -C_{ik \eta \rm RX} +C_{ij \eta \rm TX} +C_{ik \eta \rm TX}  \right) \nonumber\\
  &+& \frac{\sdws/2}{2\,\sme+3\,\sse+2\,\sdws} \left( -C_{ij \eta \rm RX} +C_{ik \eta \rm RX} +C_{ij \eta \rm TX} -C_{ik \eta \rm TX}  \right) \nonumber\\
  &+& \frac{13\,\sdws/2}{26\,\sme+25\,\sse+26\,\sdws} \left( C_{ij \eta \rm RX} +C_{ik \eta \rm RX} +C_{ij \eta \rm TX} +C_{ik \eta \rm TX}  \right) \nonumber\\
  &+& \frac{13\,\sdws/2}{26\,\sme+27\,\sse+26\,\sdws} \left( C_{ij \eta \rm RX} -C_{ik \eta \rm RX} +C_{ij \eta \rm TX} -C_{ik \eta \rm TX}  \right) \nonumber \\
  \,,
  \label{eq:dC_etaTX}\\ \hspace*{-\mathindent}
  \lim_{N\rightarrow\infty} \delta \widehat{C}^\mathrm{LS}_{ij \varphi \rm RX} 
  &=& \frac{\sdws/4}{\smp+\sdws} \left( C_{ij \varphi \rm RX} +C_{ik \varphi \rm RX} -C_{ij \varphi \rm TX} -C_{ik \varphi \rm TX}  \right) \nonumber\\
  &+& \frac{\sdws/4}{\smp+2\,\ssp+\sdws} \left( C_{ij \varphi \rm RX} -C_{ik \varphi \rm RX} -C_{ij \varphi \rm TX} +C_{ik \varphi \rm TX}  \right) \nonumber\\
  &+& \frac{13\,\sdws/4}{13\,\smp+12\,\ssp+13\,\sdws} \left( C_{ij \varphi \rm RX} +C_{ik \varphi \rm RX} +C_{ij \varphi \rm TX} +C_{ik \varphi \rm TX}  \right) \nonumber\\
  &+& \frac{13\,\sdws/4}{13\,\smp+14\,\ssp+13\,\sdws} \left( C_{ij \varphi \rm RX} -C_{ik \varphi \rm RX} +C_{ij \varphi \rm TX} -C_{ik \varphi \rm TX}  \right) \nonumber \\ 
  \,,
  \label{eq:dC_phiRX}\\ \hspace*{-\mathindent}
  \lim_{N\rightarrow\infty} \delta \widehat{C}^\mathrm{LS}_{ij \varphi \rm TX} 
  &=& \frac{\sdws/4}{\smp+\sdws} \left( -C_{ij \varphi \rm RX} -C_{ik \varphi \rm RX} +C_{ij \varphi \rm TX} +C_{ik \varphi \rm TX}  \right) \nonumber\\
  &+& \frac{\sdws/4}{\smp+2\,\ssp+\sdws} \left( -C_{ij \varphi \rm RX} +C_{ik \varphi \rm RX} +C_{ij \varphi \rm TX} -C_{ik \varphi \rm TX}  \right) \nonumber\\
  &+& \frac{13\,\sdws/4}{13\,\smp+12\,\ssp+13\,\sdws} \left( C_{ij \varphi \rm RX} +C_{ik \varphi \rm RX} +C_{ij \varphi \rm TX} +C_{ik \varphi \rm TX}  \right) \nonumber\\
  &+& \frac{13\,\sdws/4}{13\,\smp+14\,\ssp+13\,\sdws} \left( C_{ij \varphi \rm RX} -C_{ik \varphi \rm RX} +C_{ij \varphi \rm TX} -C_{ik \varphi \rm TX}  \right) \nonumber \\
  \,,
  \label{eq:dC_phiTX}
\end{eqnarray}
with $i,j,k\in\{1,2,3\}$ distinct, and 
\begin{eqnarray}
  \sigma_\alpha^\mathrm{2B} = \frac{1}{N} \sum_{n=1}^N \alpha^\mathrm{B}_{ij}(t_n)^2 \,,
  \qquad &\forall i,j,\ \alpha\in\{\eta,\varphi\}\,,\ 
  \mathrm{B}\in\{\mathrm{SC},\mathrm{MO}\} \,, \label{eq:sigma_jitter}\\
  \sigma^\mathrm{2DWS} = \frac{1}{N} \sum_{n=1}^N \delta\alpha(t_n)^2 \,,
  \qquad &\alpha\in\{\eta,\varphi\} \,.
\end{eqnarray}
If the mean jitter angles $\overline{\alpha}_{ij}^\mathrm{B}= \frac{1}{N} \sum_{n=1}^N \alpha^\mathrm{B}_{ij}(t_n)$ are not zero, we would have to subtract these mean values from the $\alpha^\mathrm{B}_{ij}(t_n)$ terms in equation~\eref{eq:sigma_jitter}. We assume that the readout noise has a mean of zero.
Equations~\eref{eq:dC_etaRX}-\eref{eq:dC_phiTX} show that the bias of the LS estimates for non-zero readout noise depends on the coupling coefficients and the variances of the readout noise and the jitter.
Note that in LISA we only have access to the combined measurement of SC and MOSA jitter. Our analysis assumes prior knowledge of their amplitude and frequency shapes. However, the same analysis can be performed using the unsplit angular DWS readout and distinguishing between the measurements of the different interferometers.

For colored jitter, the presented equations become more complex due to autocorrelations between angle measurements and their delayed versions are non-zero. We present the corresponding equations in the \ref{app:bias}.
We assume that the readout error $\delta\alpha$ is normally distributed in the frequency range of interest and is the same for all measurements.
Note also that we assume that all propagation times between the SC are different but constant. In fact, the arm length will change by up to 0.035\% in a day \cite{Paczkowski2022}, which corresponds to a propagation time difference of 3\,ms.
Equal or time-varying delays will change the equations slightly, as the correlations between the jitter measurements will differ. 

\subsection{Instrumental variables}
\label{sec:TTL_estimators_IV}

The IV estimator is defined as the LS estimator with the only difference that we multiply the measurements of two different time stamps, i.e.\ we shift the index of the time stamps of one of the multipliers by one \cite[Ch.~1,\,7]{Pintelon2001}. 
We have
\begin{eqnarray}
  \widehat{\mathbf{C}}^\mathrm{IV}
  = \left(\sum_{n=1}^{3N-1} \left[\bbeta_\mathrm{TDI}^\mathrm{DWS}(t_{n+1})\right]^T\cdot\balpha_\mathrm{TDI}^\mathrm{DWS}(t_n)\right)^{-1}
  \cdot \left(\sum_{n=1}^{3N-1} \left[\bbeta_\mathrm{TDI}^\mathrm{DWS}(t_{n+1})\right]^T\cdot \mathbf{S}_\mathrm{TDI}^\mathrm{IFO}(t_n)\right) \nonumber \\
  \label{eq:IV}
\end{eqnarray}
with 
\begin{description}
  \item[$\bbeta_\mathrm{TDI}^\mathrm{DWS}(t_n)\in\mathbb{R}^{1\times24}$ :] It is $\bbeta_\mathrm{TDI}^\mathrm{DWS}(t_n)=\cases{0 & for $n \mod N = 1$ \\ \balpha_\mathrm{TDI}^\mathrm{DWS}(t_n) & otherwise\\}$.
  We add the zeros to the vector to avoid multiplying the TDI\,X entries of the measurement at time $t_N$ with the TDI\,Y entries of the measurement at time $t_1$, and similarly for TDI\,Y and TDI\,Z.
\end{description}

The variance of the IV estimator is larger than that of the LS estimator. However, it is unbiased in most cases. Therefore, the IV method will perform better than the LS estimator if the angular readout is noisy and the data set is sufficiently large \cite[Ch.~1,\,7]{Pintelon2001}. 

The convergence of the IV method depends on the convergence of the denominator of equation~\eref{eq:IV}:
\begin{eqnarray}
  &\lim_{N\rightarrow\infty} \sum_{n=1}^{3N-1} \left[\bbeta_\mathrm{TDI}^\mathrm{DWS}(t_{n+1})\right]^T\cdot\balpha_\mathrm{TDI}^\mathrm{DWS}(t_n) \nonumber\\
  & = \lim_{N\rightarrow\infty} \sum_{n=1}^{3N-1} \left[ \bbeta_\mathrm{TDI}^\mathrm{SC}(t_{n+1})\right]^T\cdot\balpha_\mathrm{TDI}^\mathrm{SC}(t_n)  
  + \left[ \bbeta_\mathrm{TDI}^\mathrm{MO}(t_{n+1})\right]^T\cdot\balpha_\mathrm{TDI}^\mathrm{MO}(t_n) \nonumber\\
  & \cases{=0 & for white SC \textbf{and} MOSA jitter \\
  \neq0 & for colored SC \textbf{or} MOSA jitter}
\end{eqnarray}
with $\balpha_\mathrm{TDI}^\mathrm{SC/MOSA}$ the SC or MOSA jitter after TDI. $\bbeta_\mathrm{TDI}^\mathrm{SC/MOSA}$ is defined in the same way as above.
We assume that the SC and MOSA jitter are uncorrelated with each other and with the readout noise. 
If the SC or MOSA jitter is colored, the IV method converges to the true, i.e.\ unbiased, TTL coupling coefficient.
Note that the IV method may converge more slowly as the DWS readout noise increases. Therefore, when dealing with finite data sets, we potentially see a small increase in the estimation error as the DWS readout noise goes up. However, this is not a bias of the estimator itself.

\section{Performance of the estimators in simulations}
\label{sec:TTL_uncertainty}

In this section, we will show the performance of the two estimators introduced in \sref{sec:TTL_estimators} for three different jitter cases.
First, we assume the jitter settings comparable to \cite{Paczkowski2022}. In the frequency range of interest, this jitter is white.
Second, we look at the more realistic case of colored jitter based on \cite{George2022}.
Third, we investigate the performance for maneuvers, i.e.\ short periods of intentionally increased SC and MOSA jitter. We used the settings suggested in \cite{Wegener2024}.
The amplitude spectral densities (ASDs) of the measured jitters, i.e.\ the true jitters plus the readout noise (see below), are shown in \fref{fig:jitter_asd}.

\begin{figure}
\begin{indented}
\item[]\includegraphics[width=0.45\textwidth]{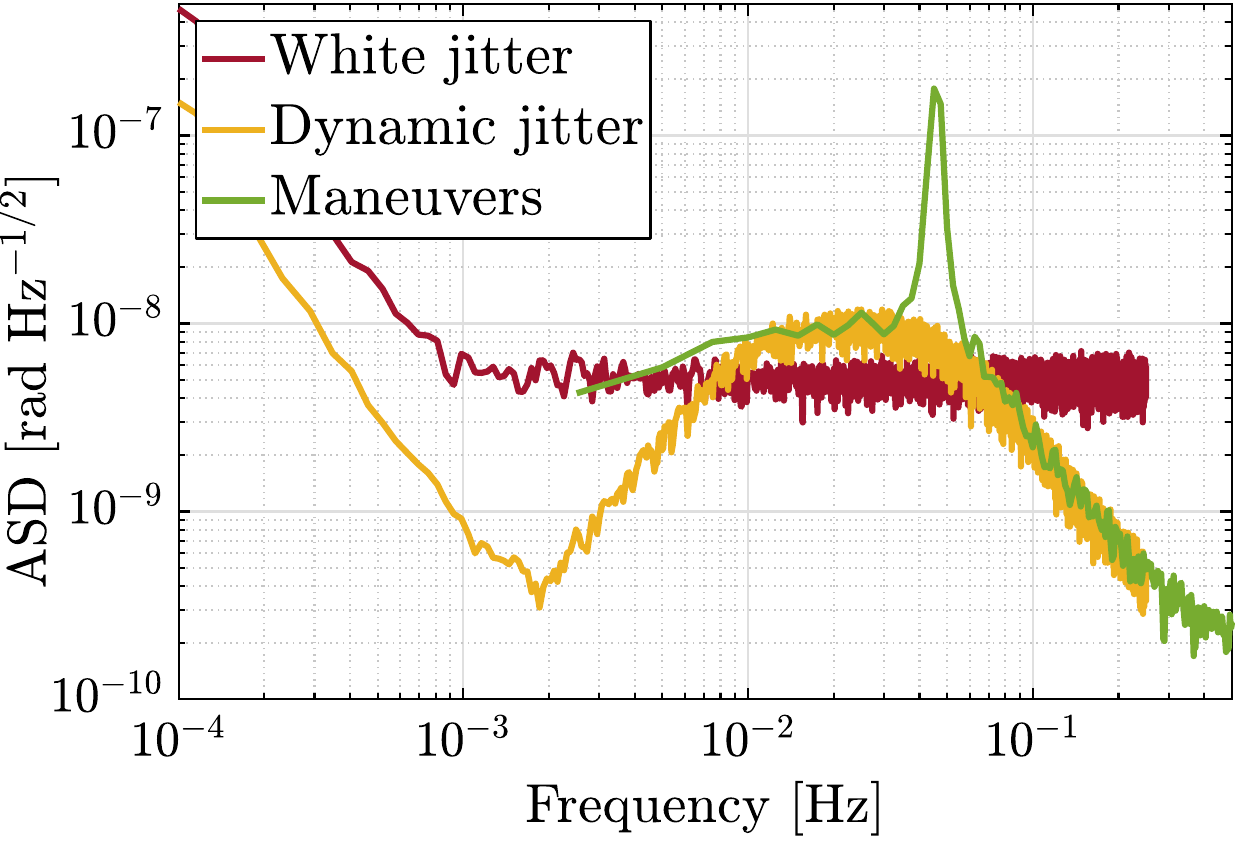}\hfill
  \includegraphics[width=0.45\textwidth]{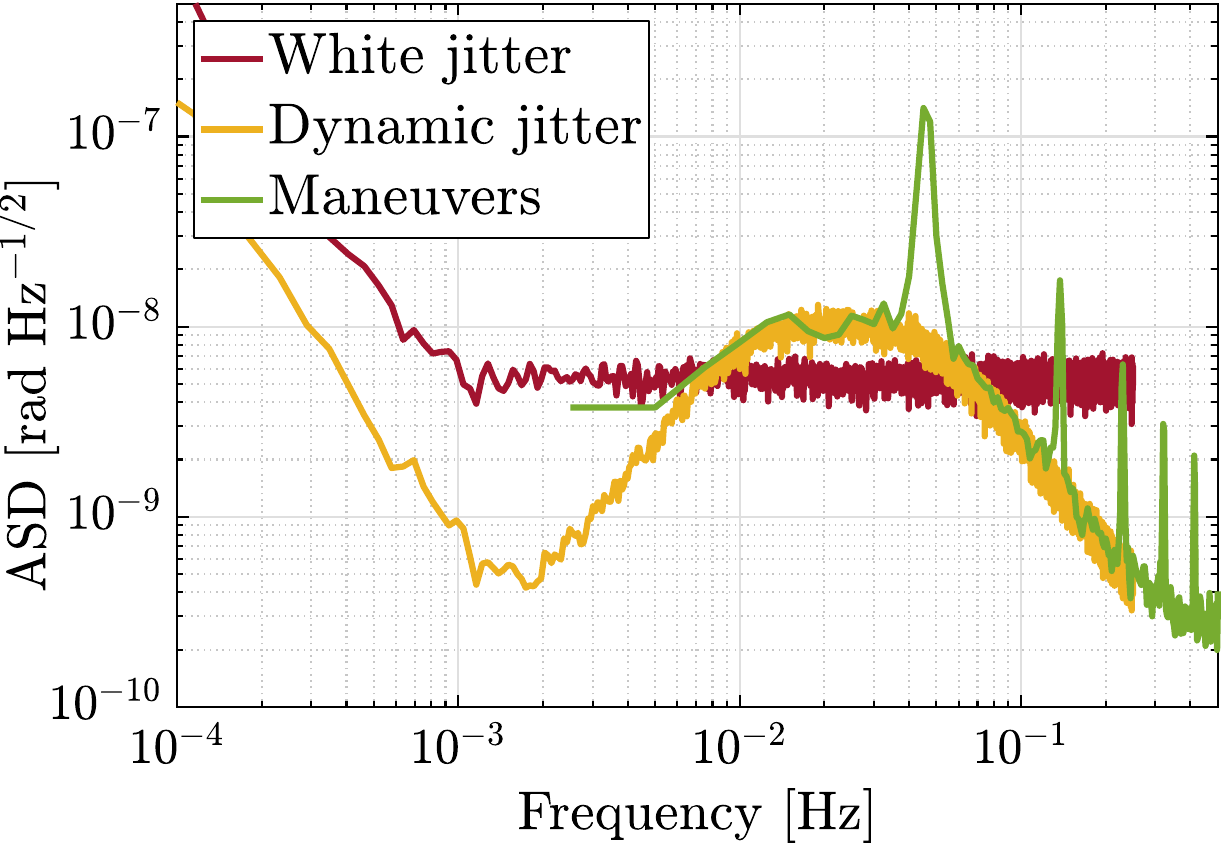}
  \caption{ASDs of the jitters that we consider in our performance analysis. Left: Measured pitch jitter $\eta^{\rm DWS}$. Right: Measured yaw jitter $\varphi^{\rm DWS}$. The pitch and yaw jitter mechanisms are different in the maneuver scenario, see \sref{sec:performance_maneuvers}.}
  \label{fig:jitter_asd}
\end{indented}
\end{figure}

\subsection{Simulation details}
\label{sec:TTL_simulation}

All our simulations were performed in MATLAB. 
First, we generated random realizations of the single link measurements for the instrument noises and angular jitters from their ASDs. 
In the first two scenarios, we simulated one day of data with a sampling frequency of 0.5\,Hz. This is motivated by the current plan to estimate the TTL coupling coefficients as part of the daily instrument monitoring. 
The maneuvers would be performed in a much shorter time. We generate data for 2000\,s, which is sufficiently long to cover the maneuver time, and a sampling frequency of 4\,Hz.
In all three investigated cases, the ASD of the nominal DWS readout noise is given by \cite{Paczkowski2022}
\begin{eqnarray}
  S^{\mathrm{DWS},1/2} = \frac{70}{335}\,\mathrm{nrad}/\sqrt{\mathrm{Hz}}\
  \sqrt{ 1+ \left(\frac{2\,\mathrm{mHz}}{f}\right)^4 } \,,
  \label{eq:DWS}
\end{eqnarray}
where 335 is the magnification factor of the telescope and imaging systems along the beam path.
The ASDs of the considered jitters will be defined below.
For the ASDs of the instrument noises in the single link length measurements compare \sref{app:ifo_noise}.
Moreover, we set all TTL coupling coefficients to 2.3\,mm/rad as in \cite{Paczkowski2022}.

To further prepare the data for coefficient estimation, we computed the delayed version of the single link measurements and filtered the data strings.
Only data above 2\,mHz are considered. In this frequency range, TTL coupling is assumed to be dominant. 
We also cut off frequencies above 0.9 times the sampling frequency, where unwanted filter or aliasing effects might add to the data.
In addition, a few hundred data points are truncated at the beginning of the data trunk to compensate for filter effects. 
Then we computed the 2nd generation TDI Michelson combinations of the length and the angular measurements. 
These were then substituted into the estimators presented in \sref{sec:TTL_estimators}.
We repeated the coefficient estimation for 100 data sets to average out the influence of the individual data realizations.

\subsection{White jitter}
\label{sec:performance_white}

\begin{figure}
\begin{indented}
\item[]\includegraphics[width=0.45\textwidth]{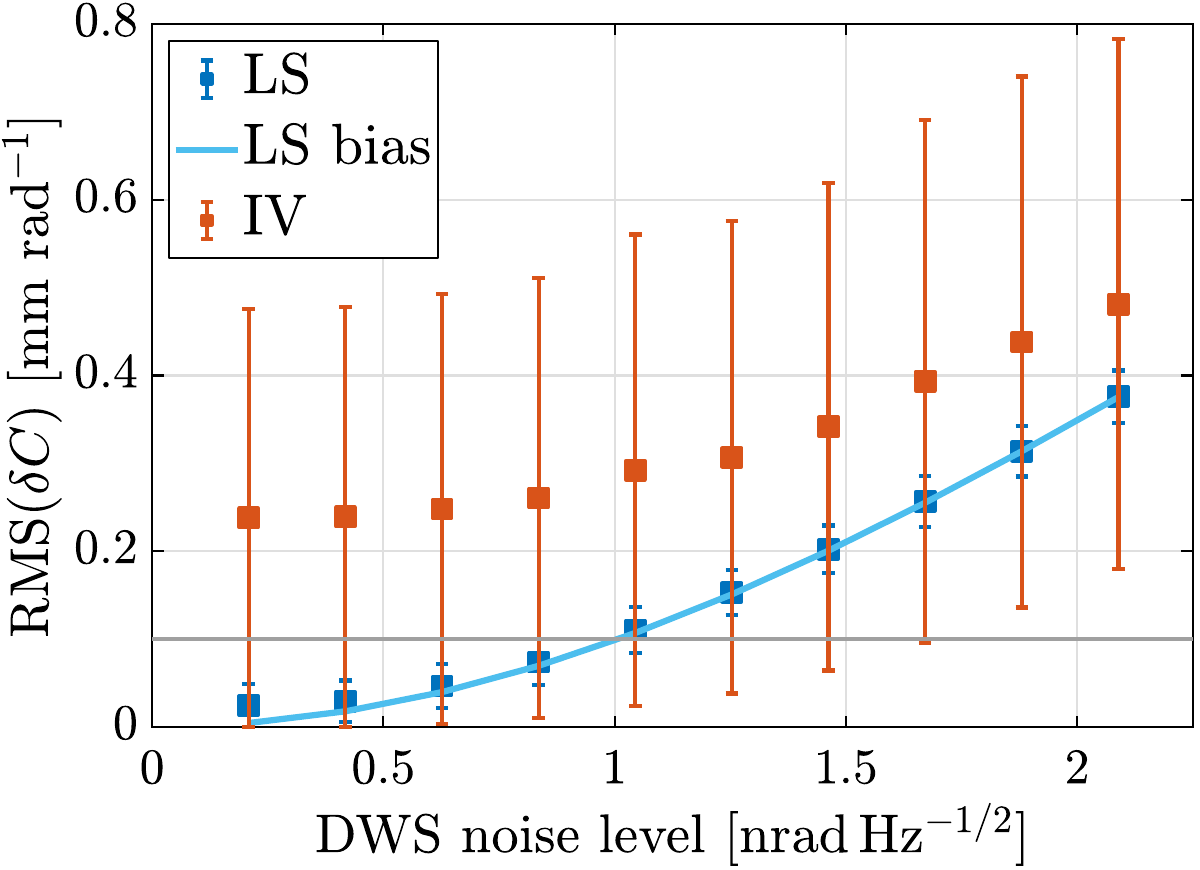}
  \caption{Root mean square (RMS) of the coefficient estimation error $\delta C$ for increased DWS readout noise levels in the case of white jitter. The ten data points correspond to DWS noise levels increased by factors of one to ten. Each RMS value was computed considering 100 1-day data sets with different noise and jitter realizations. The error bars are the computed standard deviations. The cyan line shows the analytically computed bias of the LS estimator. The gray line highlights the 0.1\,mm/rad deviation level.}
  \label{fig:RMS_white}
\end{indented}
\end{figure}

We begin with investigating the estimator performance for SC and MOSA jitter shapes based on \cite{Paczkowski2022}, which are white above 2\,mHz. These jitter levels hold as requirements for the LISA mission.
\begin{eqnarray}
  S^{\mathrm{SC},1/2}_\alpha &= 5\,\mathrm{nrad}/\sqrt{\mathrm{Hz}}\
  \sqrt{ 1+ \left(\frac{0.8\,\mathrm{mHz}}{f}\right)^4 }\,, \qquad \alpha\in\{\varphi,\eta,\theta\}\,, \label{eq:SC_SP}\\
  S^{\mathrm{MOSA},1/2}_\varphi &= 2\,\mathrm{nrad}/\sqrt{\mathrm{Hz}}\
  \sqrt{ 1+ \left(\frac{0.8\,\mathrm{mHz}}{f}\right)^4 }\,, \label{eq:MOSA_phi_SP}\\
  S^{\mathrm{MOSA},1/2}_\eta &= 1\,\mathrm{nrad}/\sqrt{\mathrm{Hz}}\
  \sqrt{ 1+ \left(\frac{0.8\,\mathrm{mHz}}{f}\right)^4 }\,. \label{eq:MOSA_eta_SP}
\end{eqnarray}
Note that we stated above that the IV estimator does not converge for white jitter.
This is clearly visible in \fref{fig:RMS_white}. The errors and their uncertainties of the IV results are much larger than for the LS estimator. The positive slope of the mean IV estimation errors is related to the DWS increase. However, this is not a systematic bias of the IV estimator, but would diminish for longer data sets as we have discussed in \sref{sec:TTL_estimators_IV}.
The LS estimator shows a clear bias for increasing DWS readout noise levels. The bias agrees well with the analytically calculated bias using equations~\eref{eq:dC_etaRX}-\eref{eq:dC_phiTX} (cyan curve). 
For low DWS noise levels, the LS estimator is mostly limited by the noise in the longitudinal readout. This explains the small discrepancies between the estimator results and the computed bias for these DWS readout noise levels.
Our results for the LS estimator are comparable to the coefficient errors shown in~\cite[cf.\ fig.\,15]{Paczkowski2022}, where a MCMC fitting algorithms was used.

\begin{figure}
\begin{indented}
\item[]\includegraphics[width=0.45\textwidth]{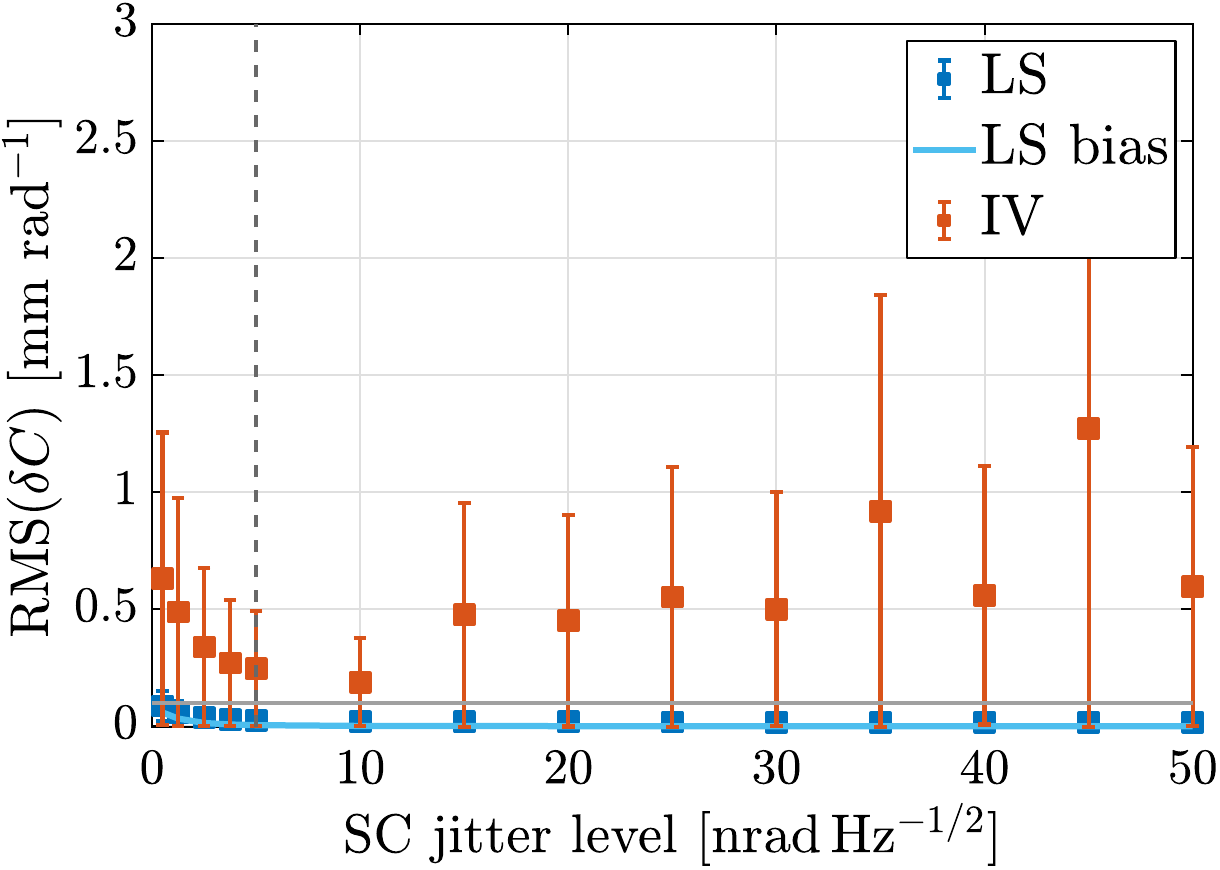}\hfill
  \includegraphics[width=0.45\textwidth]{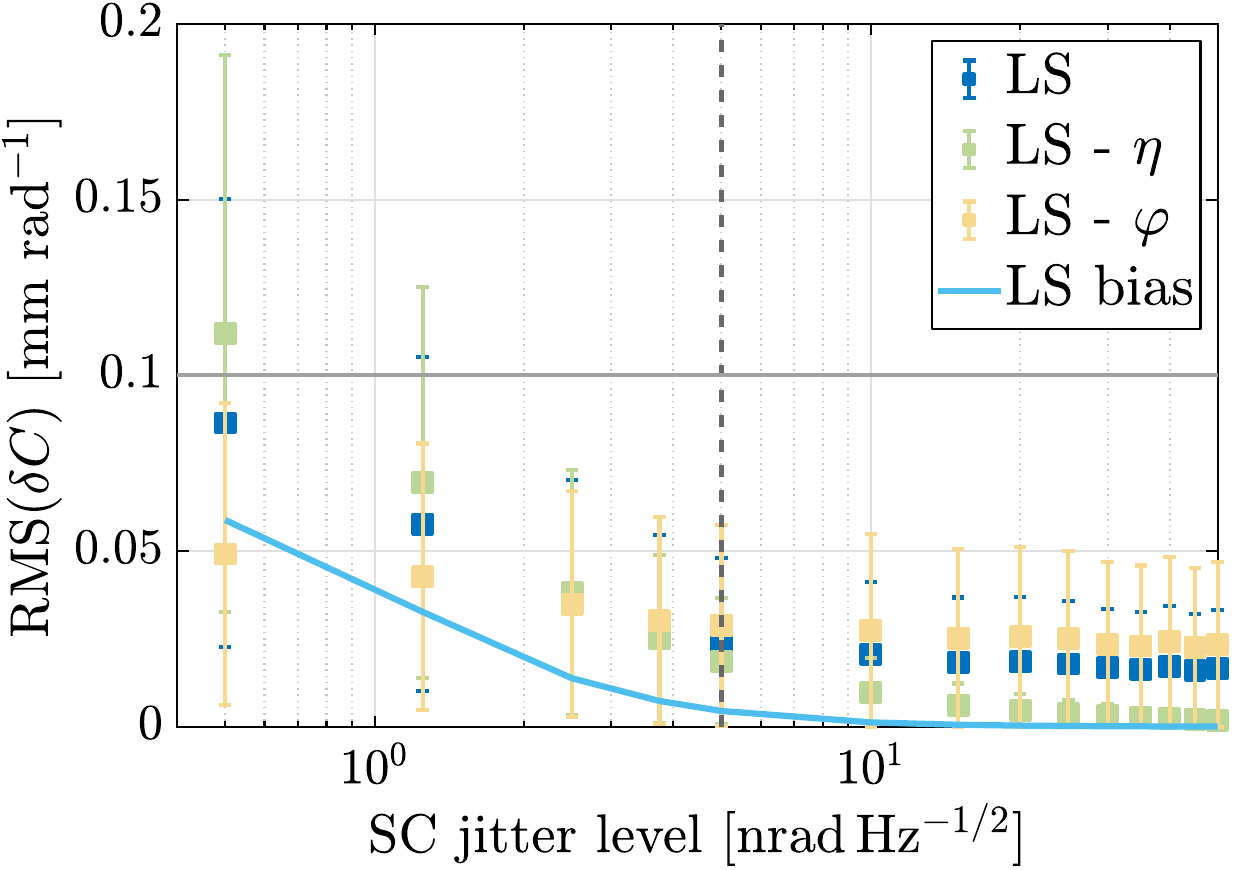}
  \caption{RMS of the coefficient estimation error $\delta C$ for different SC jitter levels in the case of white jitter. 
  Left: RMS of the estimation errors of the LS and the IV method.
  Right: RMS of the estimation errors of the pitch ($\eta$) and yaw ($\varphi$) coefficients computed with the LS method.
  This graph has a logarithmic $x$-axis to better show the estimation errors for small SC jitter levels.
  Each RMS value was computed considering 100 1-day data sets with different noise and jitter realizations. The error bars are the computed standard deviations. 
  The cyan line shows the computed bias of the LS estimator. The gray dashed vertical line marks the nominal SC jitter level. The gray horizontal line highlights the 0.1\,mm/rad deviation level.}
  \label{fig:RMS_white_jitter}
\end{indented}
\end{figure}

Next, we looked at the estimator performance for different SC jitter levels. We kept the MOSA jitter levels at their nominal levels, see equations~\eref{eq:MOSA_phi_SP} and \eref{eq:MOSA_eta_SP}.
The results are shown in \fref{fig:RMS_white_jitter}.
We see again that the IV method is very inaccurate for white jitter. 
The LS estimates deteriorate for smaller jitter amplitudes, but do not improve significantly for large jitters. This is especially true for the yaw coefficients, see the yellow markers in the right figure. 
When the SC jitter is the dominant jitter contribution, the yaw jitter in both inter-satellite interferometers on the same SC cannot be well distinguished and the associated coupling coefficients become highly correlated (see \ref{app:correlation}).
This correlation is not captured by the analytical bias equation, which therefore deviates from the plotted RMS values. 
However, we see that it is consistent with the RMSs of the pitch estimation errors for large SC jitter, which can be better distinguished.
For small jitter levels, the instrument noise adds considerably to the estimates. 
Thus, the computed bias can explain only a part of the estimation error increase for reduced SC jitter levels. 

\subsection{Colored jitter}
\label{sec:performance_dynamic}

\begin{figure}
\begin{indented}
\item[]\includegraphics[width=0.45\textwidth]{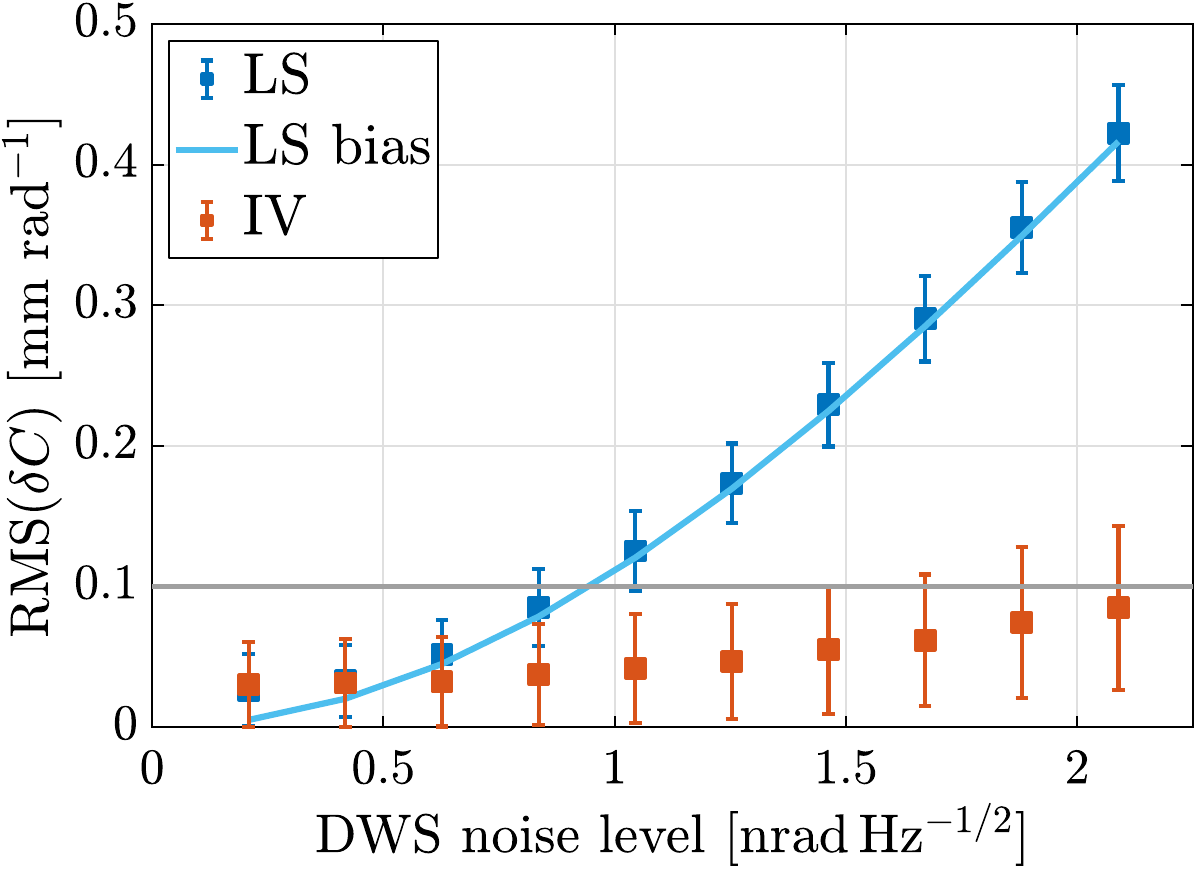}
  \caption{RMS of the coefficient estimation error $\delta C$ for increased DWS readout noise levels in the case of colored jitter. The ten data points correspond to DWS noise levels increased by factors of one to ten. Each RMS value was computed considering 100 1-day data sets with different noise and jitter realizations. The error bars are the computed standard deviations. The cyan line shows the computed bias of the LS estimator. The gray line highlights the 0.1\,mm/rad deviation level.}
  \label{fig:RMS_dynamic}
\end{indented}
\end{figure}

While the jitter definition in the previous case set a requirement for maximum jitter, these shapes are not very realistic. In fact, we expect the jitter to roll off at high and low frequencies \cite{Inchauspe2022}. Therefore, we consider here a colored SC jitter similar to~\cite{George2022}.  
\begin{eqnarray} \hspace*{-\mathindent}
  S^{\mathrm{SC},1/2}_\alpha &= 0.7^2\cdot 40\,\mathrm{frad}/\sqrt{\mathrm{Hz}}\
  \left[1+\left(\frac{f}{20\,\mu\mathrm{Hz}}\right)^2\right]
  \left\lbrace \frac{(10\,\mathrm{mHz})^2}{\sqrt{0.7^2 \left[f^2-(10\,\mathrm{mHz})^2\right]^2+(10\,\mathrm{mHz})^2 f^2}} \right\rbrace 
  \nonumber\\
  &\quad \cdot \left\lbrace \frac{(50\,\mathrm{mHz})^2}{\sqrt{0.7^2 \left[f^2-(50\,\mathrm{mHz})^2\right]^2+(50\,\mathrm{mHz})^2 f^2}} \right\rbrace
  \left/ \left[1+\left(\frac{f}{8\,\mathrm{Hz}}\right)^2\right] \right. \,.
\end{eqnarray}
for $\alpha\in\{\varphi,\eta,\theta\}$.
The definition of MOSA jitter in \cite{George2022} is outdated, but to the authors' knowledge, no updated realistic MOSA jitter shape has been published. Therefore, we decided to assume the same jitter shape for the $\varphi$-MOSA and the SC in this scenario. The ratio of their magnitudes is the same as in the previously studied case.
\begin{eqnarray}
  S^{\mathrm{MOSA},1/2}_\varphi &= 0.4 \cdot S^{\mathrm{SC},1/2}_\alpha \,, \\
  S^{\mathrm{MOSA},1/2}_\eta &= 0 \,.
\end{eqnarray}
The performance of the estimators is shown in \fref{fig:RMS_dynamic}.
As in the case of white jitter, the LS estimator is biased, but this bias can be calculated using our analytical equation.
The IV estimator performs much better in this scenario. The coefficient estimation error stays below 0.1\,mm/rad even for a tenfold increase in DWS readout noise (equation~\eref{eq:DWS}).

\begin{figure}
\begin{indented}
\item[]\includegraphics[width=0.45\textwidth]{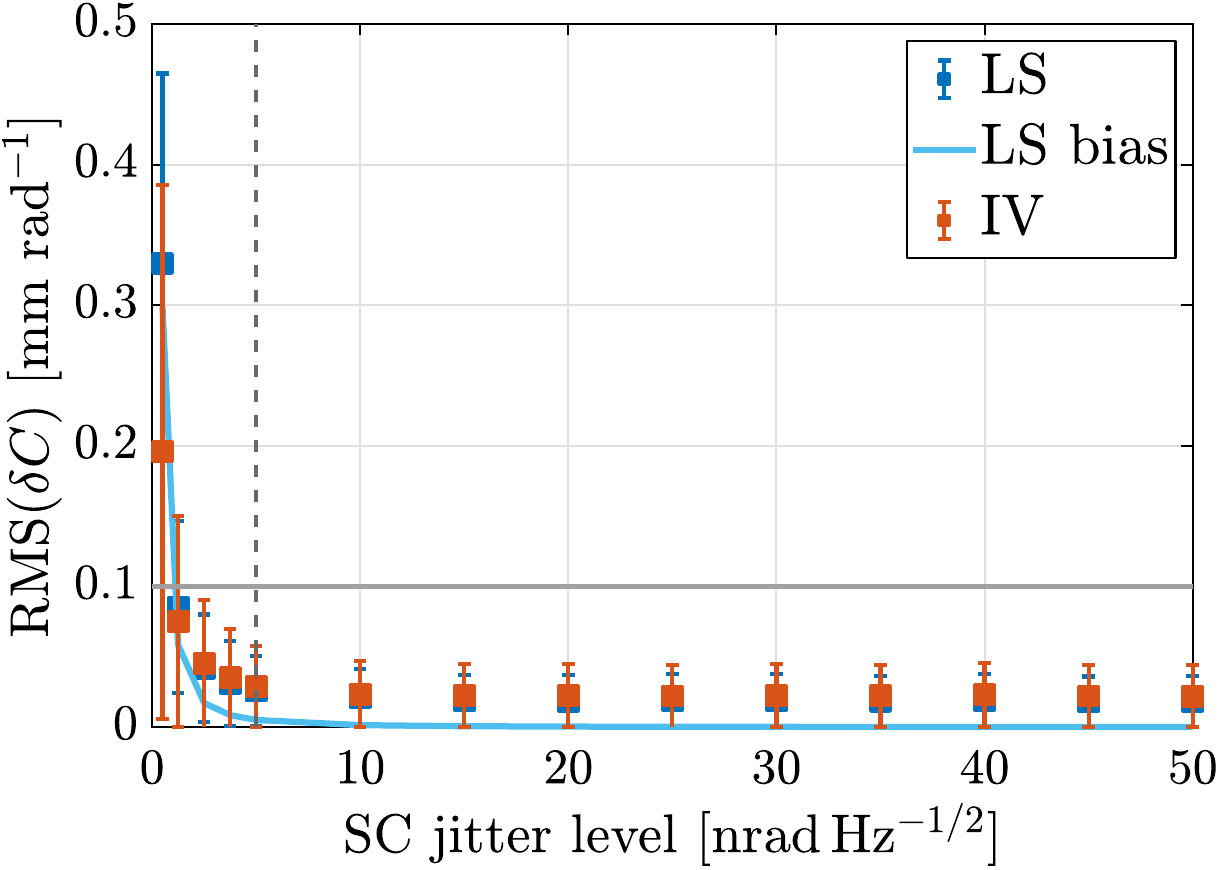}\hfill
  \includegraphics[width=0.45\textwidth]{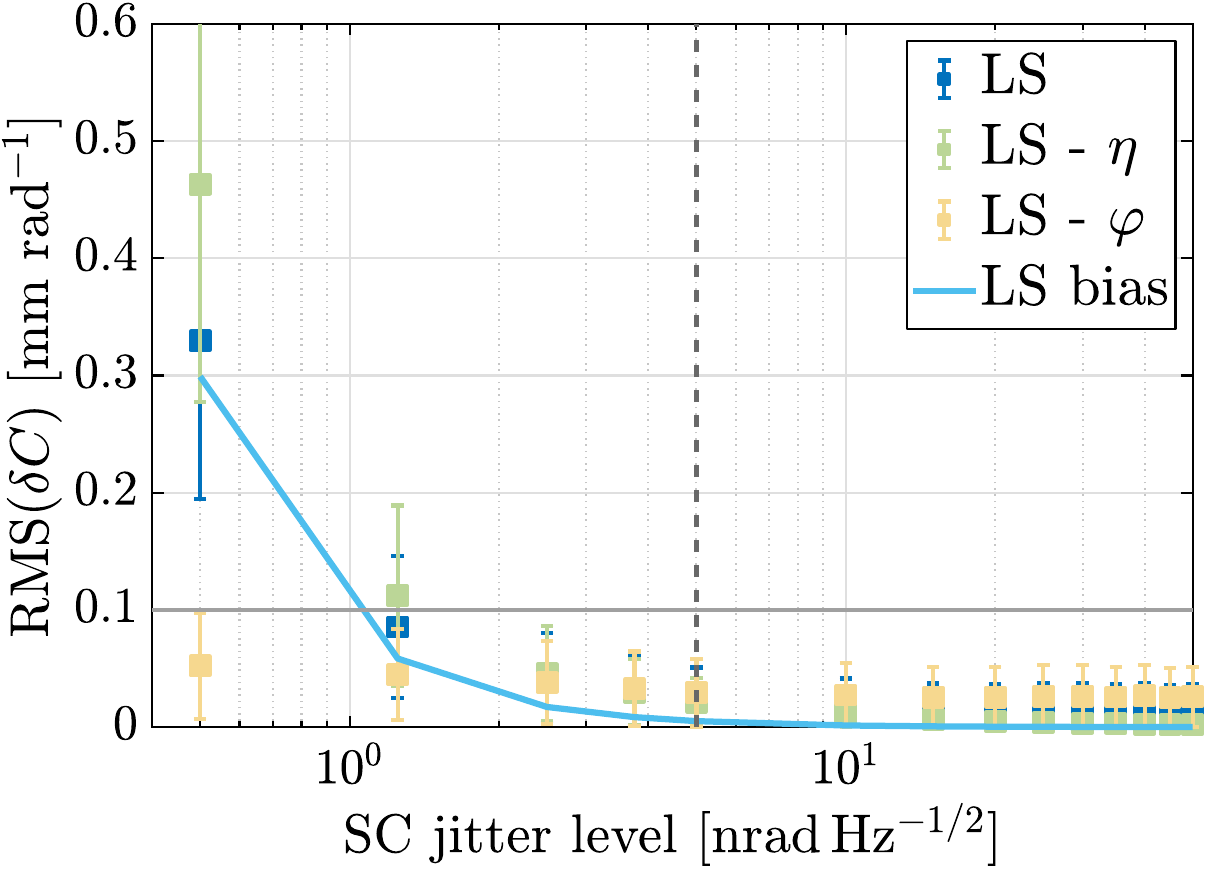}
  \caption{RMS of the coefficient estimation error $\delta C$ for different SC jitter levels in the case of colored jitter. 
  Left: RMS of the estimation errors of the LS and the IV method.
  Right: RMS of the estimation errors of the pitch ($\eta$) and yaw ($\varphi$) coefficients computed with the LS method.
  This graph has a logarithmic $x$-axis to better show the estimation errors for small SC jitter levels
  Each RMS value was computed considering 100 1-day data sets with different noise and jitter realizations. The error bars are the computed standard deviations. 
  The cyan line shows the computed bias of the LS estimator. The gray dashed vertical line marks the nominal SC jitter level. The gray horizontal line highlights the 0.1\,mm/rad deviation level.}
  \label{fig:RMS_dynamic_jitter}
\end{indented}
\end{figure}

For varying SC jitter levels, we find comparable results for the LS and IV methods, see \fref{fig:RMS_dynamic_jitter}.
As in the case of white jitter, the LS estimator does not converge to zero for large jitter levels because the yaw coefficients are highly correlated. 

\subsection{LISA Maneuvers}
\label{sec:performance_maneuvers}

LISA maneuvers for calibrating the TTL coupling are presented in the papers~\cite{Wegener2024,Houba2022a,Houba2022b}.
Here, we consider the maneuver scheme introduced in \cite{Wegener2024}.

It requires two different strategies to excite the SC and MOSAs in pitch and yaw. 
To induce distinguishable pitch jitter in all six arms, the SC is rotated in $\eta$ and $\theta$ at pairwise different frequencies or times.
In the case of TTL coupling in $\varphi$, it is not sufficient to jitter the SC since this would result in the same jitter for both MOSAs on the same SC. Instead, the MOSAs themselves are rotated step-wise. Again, this is done at different frequencies or times.
The jitter parameters are summarized in \tref{tab:maneuvers_eta} and \tref{tab:maneuvers_phi}.

\begin{table}
\caption{SC maneuvers in roll and pitch \cite{Wegener2024}. 
  These are performed in two disjunct phases of 600\,s each.
  We list the amplitudes of the sinusoidal injections in $\theta^\mathrm{SC}_i$ and $\eta^\mathrm{SC}_i$. Negative amplitudes correspond to a phase shift of $\pi$.
  The last column shows the injection frequencies.
  Since $\eta^\mathrm{SC}_i$ is always excited by the same frequency, but only by a factor $1/\sqrt{3}$ of the magnitude of $\theta^\mathrm{SC}_i$, these SC excitations add in the interferometer at one of the MOSAs and cancel in the other one.}
\label{tab:maneuvers_eta}
\begin{indented}
\item[]
\begin{tabular}{@{}ccccc}
\br
SC\,\# & Phase & Amplitude $\theta_\mathrm{SC}$ & Amplitude $\eta_\mathrm{SC}$ & Frequency\\
\mr
1  & 1 & 30\,nrad & $-$30/$\sqrt{3}$\,nrad & 43.33\,mHz\\
2  & 1 & 30\,nrad & $+$30/$\sqrt{3}$\,nrad & 41.67\,mHz\\
3  & 1 & 30\,nrad & $+$30/$\sqrt{3}$\,nrad & 43.33\,mHz\\
1  & 2 & 30\,nrad & $+$30/$\sqrt{3}$\,nrad & 43.33\,mHz\\
2  & 2 & 30\,nrad & $-$30/$\sqrt{3}$\,nrad & 43.33\,mHz\\
3  & 2 & 30\,nrad & $-$30/$\sqrt{3}$\,nrad & 41.67\,mHz\\
\br
\end{tabular}
\end{indented}
\end{table}

\begin{table}
\caption{MOSA maneuvers in yaw \cite{Wegener2024}. 
  These are performed in two disjunct phases of 600\,s each.
  We list the maximum rotation angle (`amplitudes'), the rotation step size and the period of the full cycle of steps (after maximum rotations in positive and negative direction) of the step-wise rotations in $\varphi^\mathrm{MO}_{ij}$. The periods correspond to frequencies similar to those given in \tref{tab:maneuvers_eta}.}
\label{tab:maneuvers_phi}
\begin{indented}
\item[]\begin{tabular}{@{}ccccc}
\br
MOSA\,\# & Phase & Amplitude & Step size & Period\\
\mr
12 & 2 & 30\,nrad & 1\,nrad & 21.8\,s \\
13 & 1 & 30\,nrad & 1\,nrad & 21.8\,s \\
23 & 1 & 30\,nrad & 1\,nrad & 21.8\,s \\
21 & 2 & 30\,nrad & 1\,nrad & 24.0\,s \\
31 & 1 & 30\,nrad & 1\,nrad & 24.0\,s \\
32 & 2 & 30\,nrad & 1\,nrad & 21.8\,s \\
\br
\end{tabular}
\end{indented}
\end{table}

\begin{figure}
\begin{indented}
\item[]\includegraphics[width=0.45\textwidth]{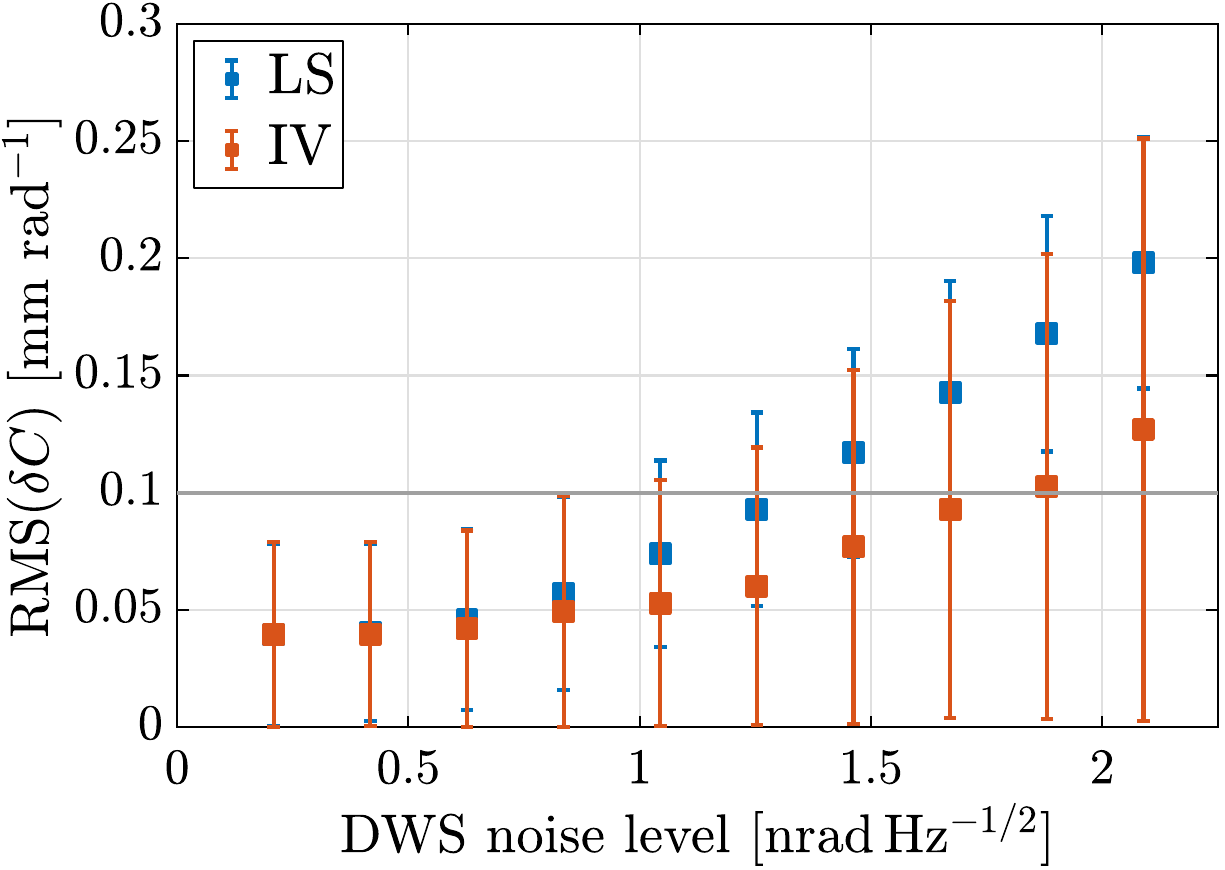}
  \caption{RMS of the coefficient estimation error $\delta C$ for increased DWS readout noise levels in the case of maneuvers. The ten data points correspond to DWS noise levels increased by factors of one to ten. Each RMS value was computed considering 100 data sets with a length of 2000\,s, a sampling frequency of 4\,Hz and different noise and jitter realizations. The error bars are the computed standard deviations. The gray line highlights the 0.1\,mm/rad deviation level.}
  \label{fig:RMS_maneuver}
\end{indented}
\end{figure}

For our analysis, we add the SC and MOSA maneuvers to the colored jitter discussed above, assuming that this would still be present if we excite the assembly.
\Fref{fig:RMS_maneuver} shows that the TTL coupling coefficient can be estimated better (LS estimator) than in the case of one day of data with colored jitter, even though we have considered less data samples than in the study cases before (data set of 2000\,s with a sampling frequency of 4\,Hz vs.\ 86400\,s with a sampling frequency of 0.5\,Hz). 
The error of the IV estimates is slightly larger than before, and their uncertainty is increased. As discussed above, this method works best for large data sets. In the case of the maneuvers, we have less than a fifth of the data points available compared to a full day of data, despite working with a higher sampling frequency. The fact that we have fewer data points in this scenario also explains why the estimates are more sensitive to the DWS readout noise (compare discussion in \fref{sec:TTL_estimators_IV}).
We do not show the analytically computed LS bias in this plot. The equations~\eref{eq:dC_etaRX_app}-\eref{eq:dC_phiTX_app} do not fully hold in the maneuver scenario since the jitters on the different SC and MOSAs are partly correlated. 

\subsection{Discussion}

In this section, we have analyzed the uncertainties of two TTL coupling coefficient estimators, the LS (equation~\eref{eq:LS}) and the IV (equation~\eref{eq:IV}) estimators, for three different types of jitter.

We find that the IV method performs almost as well as the LS algorithm for most nominal settings. In addition, it is largely insensitive to increased levels of DWS readout noise. 
On the downside, the IV method cannot be used in the case of white SC and MOSA jitters. In this scenario, the estimator does not converge. 
Since we do not expect the jitter in LISA to be white, the IV method can prove useful for the accurate estimation of TTL coupling coefficients in the presence of increased DWS readout noise.

\begin{figure}
\begin{indented}
\item[]\includegraphics[width=0.45\textwidth]{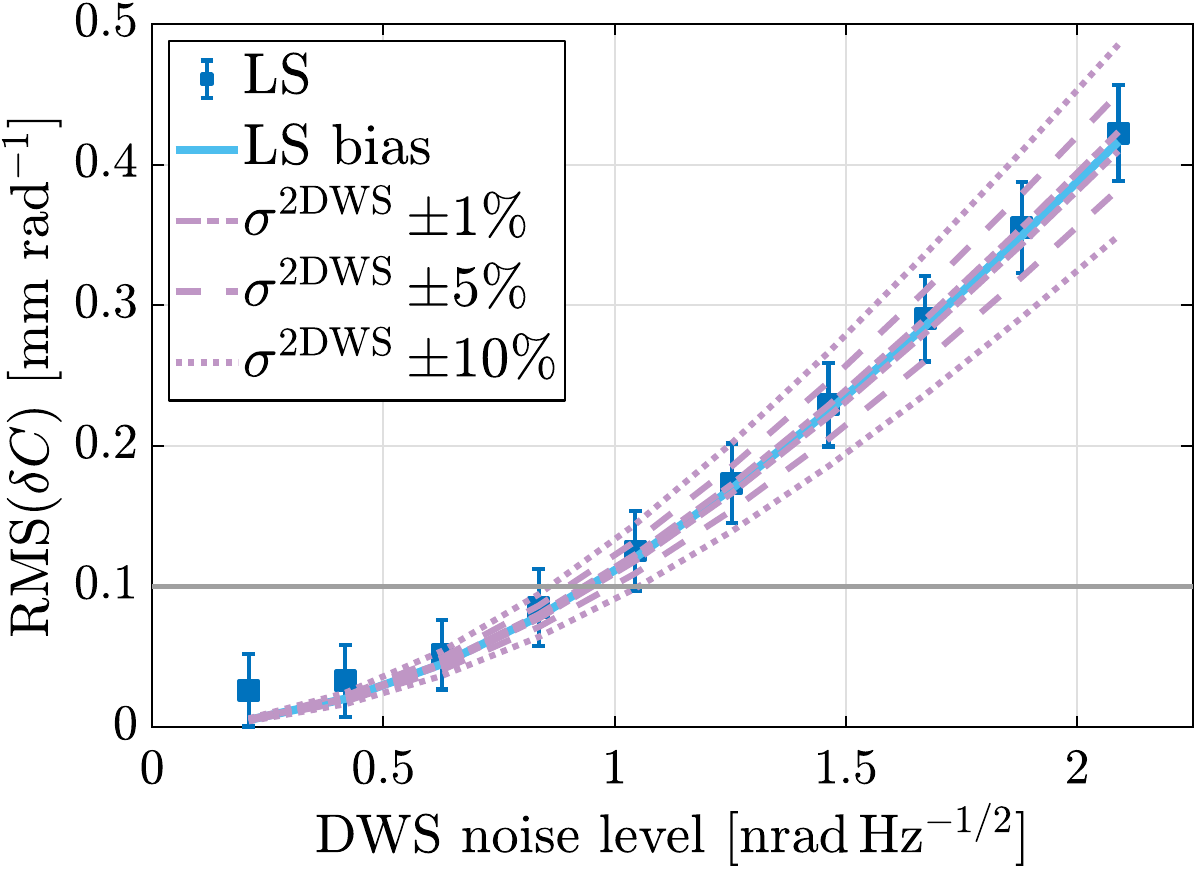}
  \caption{RMS of the LS coefficient estimation error $\delta C$ for increased DWS readout noise levels in the case of colored jitter. The ten data points correspond to DWS noise levels increased by factors of one to ten. The cyan line shows the computed bias. The purple lines show the computed bias for incorrect assumptions about the DWS readout noise level. We variate the substituted level by 1,\,5 and 10\,\%. The gray line highlights the 0.1\,mm/rad deviation level.}
  \label{fig:RMS_bias}
\end{indented}
\end{figure}

The LS estimator is biased. For all three jitter cases, we can clearly see that the LS estimator error increases with increasing DWS readout noise. 
This bias can be modeled analytically, see equations~\eref{eq:dC_etaRX}-\eref{eq:dC_phiTX} or \eref{eq:dC_etaRX_app}-\eref{eq:dC_phiTX_app}.
We find that the analytically computed bias agrees well with the estimated LS coefficient errors for white (\fref{fig:RMS_white}) and colored (\fref{fig:RMS_dynamic}) jitter. 
Therefore, the calculated bias can be used to correct the bias in the LS estimates.
To compute it, we need to know the coupling coefficients, as well as the variances and auto-correlations of the jitter and readout noise. 
In flight, we will only know the estimated, biased, and not the true coupling coefficients. However, by applying the equations iteratively, correcting the coefficient estimates at each step (i.e.\ subtracting the latest computed bias), a good approximation of the bias is obtained after a few iteration steps, see exemplary \tref{tab:bias_iteration}. %
A bigger problem may be estimating the variance of the DWS readout noise since it will be smaller than the angular jitter and has to be distinguished from the latter in the same measurement. \Fref{fig:RMS_bias} shows that the calculated bias is very sensitive to changes of $\sigma^\mathrm{2DWS}$ in the equations~\eref{eq:dC_etaRX}-\eref{eq:dC_phiTX}.
This must be taken into account when applying the equation to real data. 

\begin{table}
\caption{Comparison of the analytically computed bias of the LS estimator using either the true TTL coupling coefficients (all 2.3\,mm/rad) or the estimated coefficients (mean of the estimates of 100 realizations) as input. In the case of the estimated coefficients, we iteratively correct the estimate by the computed bias. The result for the 0.\ iteration step refers to the originally estimated coefficients.
We show the computed biases for the case of white and colored jitter case (see \sref{sec:performance_white} and \sref{sec:performance_dynamic}) and three different DWS noise levels (nominal, five times and 10 times increased).
Even for 10 times increased jitter, the analytically computed biases, using either the true or the estimated coefficients, differ by less than 0.1mm/rad. The difference decreases with each iteration step.}
\label{tab:bias_iteration}
\begin{indented}
\item[]\begin{tabular}{@{}lc p{15mm} *{5}{p{10mm}}@{\extracolsep{\fill}}}
\br
Jitter- & DWS noise level 
 & \multicolumn{5}{c}{bias (true coeff./ est.\ coeff. after \# iter.) [mm\,rad$^{-1}$]} \\
type & [nrad\,Hz$^{-1/2}$] 
 & true & 0 & 1 & 2 & 3 \\
\mr
White & 0.21 
 & 0.004 & 0.004 & 0.004 & 0.004 & 0.004 \\
 & 1.04 
 & 0.107 & 0.102 & 0.107 & 0.107 & 0.107 \\
 & 2.09 
 & 0.375 & 0.314 & 0.366 & 0.374 & 0.375 \\
 \mr
Colored & 0.21 
 & 0.005 & 0.005 & 0.005 & 0.005 & 0.005 \\
 & 1.04 
 & 0.120 & 0.114 & 0.120 & 0.120 & 0.120 \\
 & 2.09 
 & 0.416 & 0.402 & 0.413 & 0.415 & 0.415 \\
\br
\end{tabular}
\end{indented}
\end{table}

While the bias equations should hold for variations in the jitter levels, we see deviations between the LS estimates and computed bias in \fref{fig:RMS_white_jitter} and \fref{fig:RMS_dynamic_jitter}. 
For small jitter levels, the estimates are also affected by the instrument noise in the longitudinal readout. For large SC jitter levels the latter dominate the MOSA jitter. As a result, the yaw jitters on the same SC conform to each other and the corresponding coupling coefficients become highly correlated. 
Therefore, the coefficient errors for large SC jitter in these figures are mostly due to the correlations and not to the bias. 

Note that, we have set all TTL coupling coefficients to 2.3\,mm/rad in the plots shown above. This setting allowed us to easily compare the LS estimates with the computed bias.
For completeness, we show the coefficient errors for random TTL coupling coefficients in \ref{app:rand_TTL}.

\section{Performance in the presence of gravitational wave signals}
\label{sec:TTL_GW}

In this section, we investigate the effect of GWs in the estimation of the TTL coefficients by including GWs in the simulation scheme introduced in \sref{sec:TTL_simulation}. In all cases, we assume the colored SC and MOSA jitter introduced in \sref{sec:performance_dynamic}.
From the point of view of TTL estimation, GWs can be considered as a noise source in the interferometric length readout ---see~\sref{sec:TTL_estimators}, and thus, they can worsen the uncertainties of the coefficient estimates. However, they should have no effect on the bias of the coefficients since their effect on the DWS angular readout is negligible.

\begin{figure}
\begin{indented}
\item[]\includegraphics[width=0.9\textwidth]{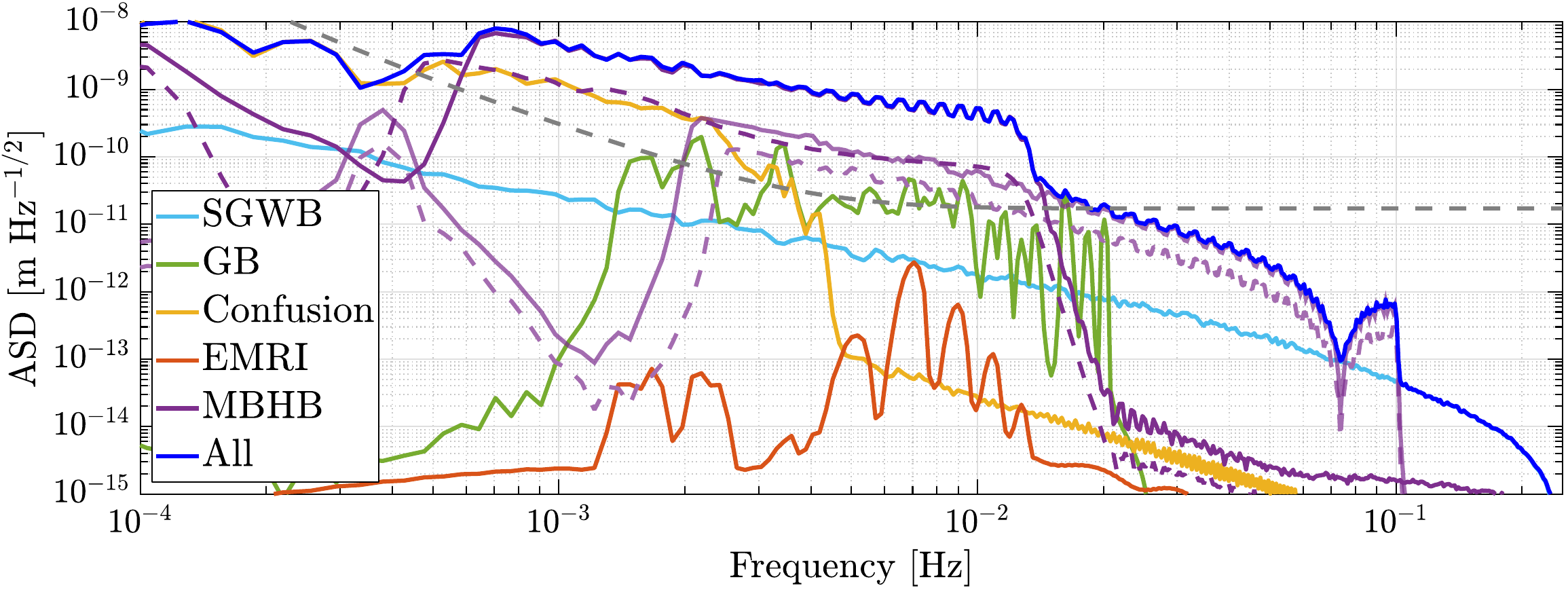} \\
\item[]\includegraphics[width=0.9\textwidth]{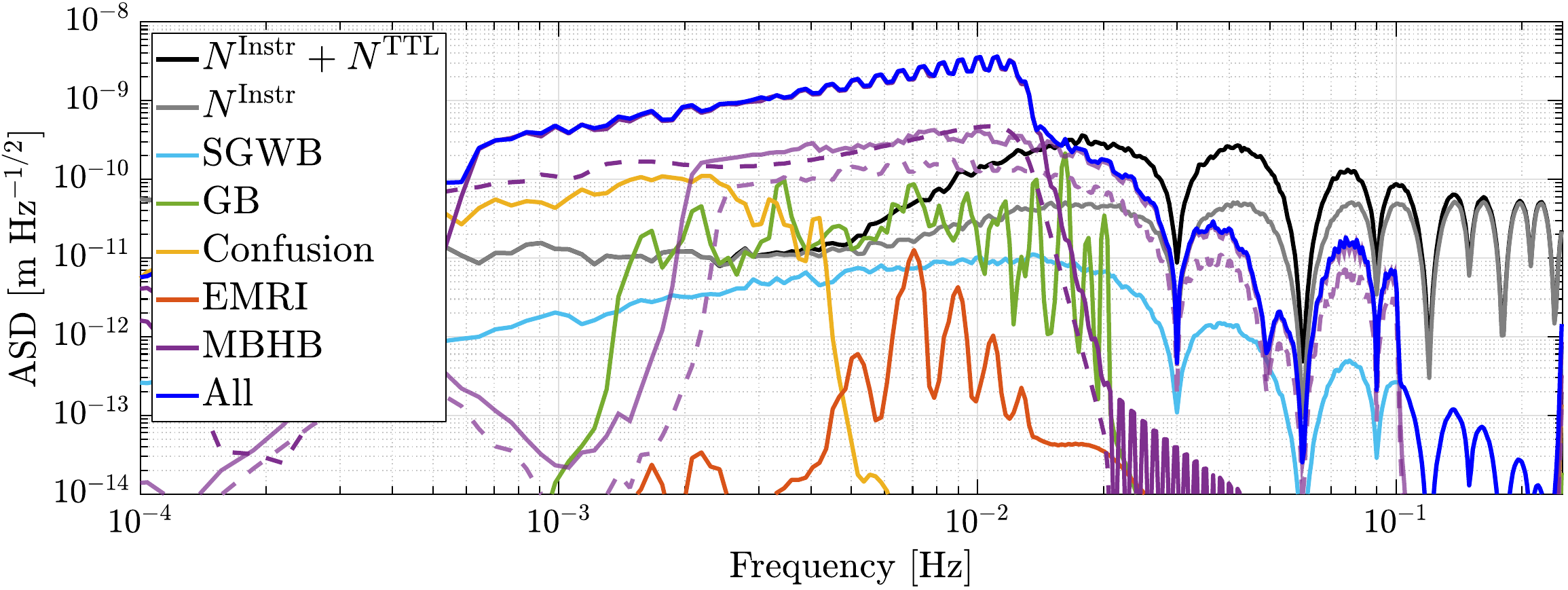} 
  \caption{ASDs of the investigated GW signals in one single link measurement (top), and the ASDs of the TDI\,X combinations of the GW responses in relation to TTL ($N^\mathrm{TTL}$) and instrument noise ($N^\mathrm{Instr}$) (bottom).
  We include four different MBHB mergers. Light purple, dashed: MBHB\,1. Light purple, solid line: MBHB\,2. Purple, dashed: MBHB\,3. Purple, solid line: MBHB\,4. Their properties are summarized in \tref{tab:gw_signals}.
  The dashed gray line in the upper plot shows the LISA requirement.
  We used the lpsd function of the MATLAB toolbox LTPDA for the computation of the PSD \cite{lpsd}.
  }
  \label{fig:ASD_gw}
\end{indented}
\end{figure}

\begin{figure}
\begin{indented}
\item[]\includegraphics[width=0.9\textwidth]{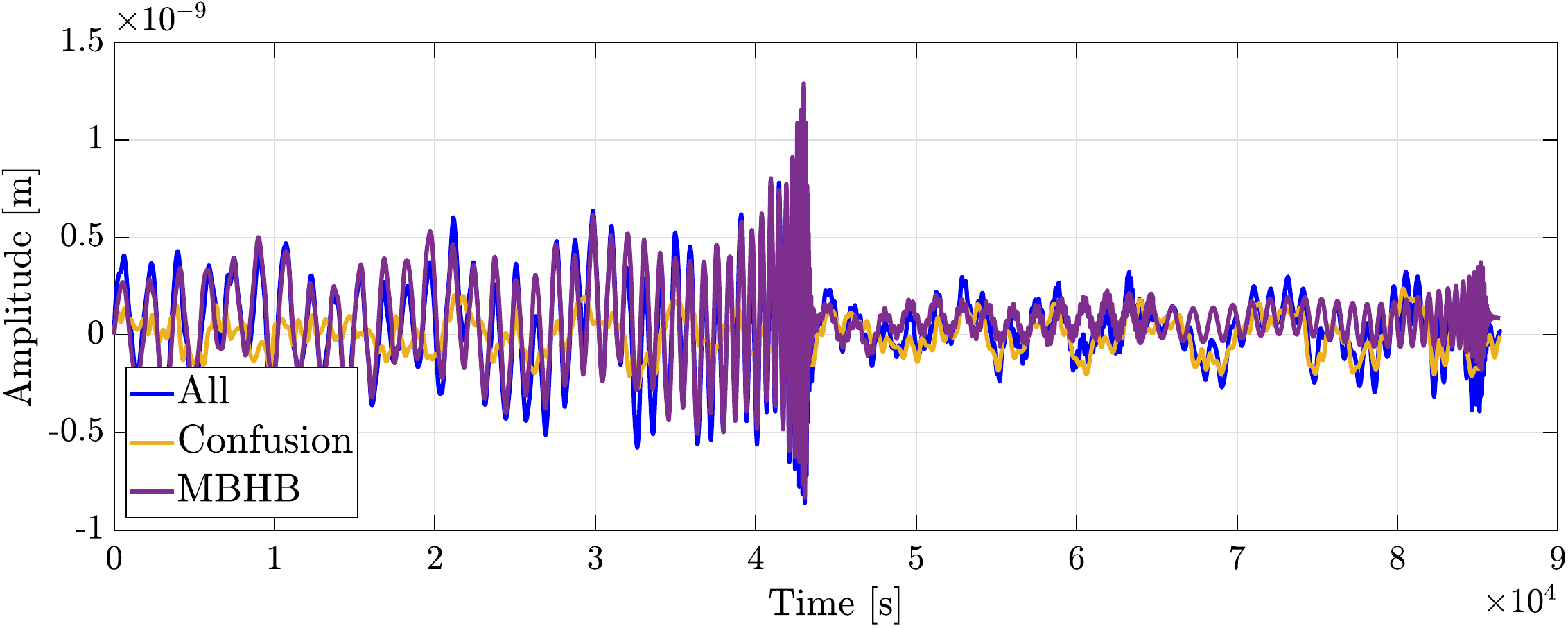} 
  \caption{Time series (single link) of the most dominant GW signal responses investigated in this manuscript, i.e.\ the confusion noise (yellow) and the MBHB mergers (purple). They account for most of the response signal in the case where all GW signals are combined (blue).
  The GW signal properties are summarized in \tref{tab:gw_signals}.
  }
  \label{fig:time_gw}
\end{indented}
\end{figure}

We selected five different sources of GWs: stochastic gravitational wave background (SGWB) from stellar-origin binary black holes (SOBBH), resolvable galactic binaries (GB), massive black hole binaries (MBHB), extreme mass ratio inspirals (EMRI), and foreground confusion noise due to unresolved galactic white dwarf binaries. We have not considered super massive black hole binaries (SMBHB) since they radiate  at frequencies well below TTL effects. For instance, a $10^{9}M_{\odot}$ black hole binary system covers a frequency range from $\sim$0.1\,nHz to $\sim$1\,$\mu$Hz~\cite{Sesana_2013}, whereas TTL coupling becomes dominant only at a few milli-hertz.
The simulated signals' ASDs compared to the LISA instrument noise are shown in~\fref{fig:ASD_gw} and summarized in \tref{tab:gw_signals}. Note that the signals in \fref{fig:ASD_gw} correspond to a measurement time of one day and, thus, some of them remain below the LISA instrument noise. We only consider one day since it is the time scale at which TTL coefficients will be recovered ---cf.~\sref{sec:TTL_estimators}. 
For convenience, we have also plotted the strongest signals in the time domain in \fref{fig:time_gw}. The figure shows the signals without instrumentation noise. 
It is clear that the galactic confusion noise and the MBHB mergers dominate the data stream, as they add up to the data showing all the GW signals together (blue line). 
In the following, details on the tools and models used for the generation of the GWs are presented. 

\begin{table}
\caption{Summary of GW signals included in the simulations.}
\label{tab:gw_signals}
\begin{indented}
\item[]
\begin{tabular}{@{}ll}
\br
     Source & Description  \\ %
     \mr
     SGWB & From SOBBHs ($h^{2}\Omega_{\rm GW}=1.15\times10^{-12}$ at 3\,mHz, $\alpha$=2/3).  \\
     GB & Resolvable ultra compact galactic binaries: \\
     & 100 loudest ones from~\cite{LISAcat}, which includes mostly white-dwarf binaries.  \\
     Confusion & Galactic white-dwarf binaries. \\
     & They are above the LISA instrument noise between $\sim$1 and $\sim$5\,mHz.  \\
     EMRI & One EMRI (23$M_{\odot}$ and $10^{6}M_{\odot}$ at 3\,Gpc).    \\
     MBHB\,1 & $10^{5}M_{\odot}$, $10^{5}M_{\odot}$ at $z$=3 (merger).  \\
     MBHB\,2 & $10^{5}M_{\odot}$, $10^{5}M_{\odot}$ at $z$=1.5 (merger). \\
     MBHB\,3 & $10^{6}M_{\odot}$, $10^{6}M_{\odot}$ at $z$=6 (merger).  \\
     MBHB\,4 & $10^{6}M_{\odot}$, $10^{6}M_{\odot}$ at $z$=2 (merger).   \\
     \br
\end{tabular}
\end{indented}
\end{table}

\subsection{Gravitational waves signals generation}

Different tools and simulators were used to generate the GWs strain in both polarizations (+ and ×) as a function of time. We used LISA GW Response~\cite{lisagw}, a Python package that computes the LISA response to gravitational waves, and contains other GW tools. The strain for the SOBBH SGWB, GBs, and foreground confusion noise were calculated using LISA GW Response, while EMRIs and MBHB mergers were generated with LISA Data Challenge (LDC) tools~\cite{ldc} using pre-existing analytical models included in the package. After adjusting the parameters (cf.~\tref{tab:gw_signals}), we utilized the source code to generate the strain of the GWs. 
Once the strain waveforms were computed, all LISA response data was generated using LISA GW Response, which outputs the LISA response as relative frequency changes (Doppler shifts). The signals were converted to length fluctuations by time-integration and scaling by the speed of light.

The LISA response to the passing GWs depends on the detector's orbit~\cite{martens2021} and its relative orientation with respect to the sources. For our purposes, we assumed unequal but constant arms, in alignment with the other data settings (cf.~\sref{sec:TTL_simulation}).
Correspondingly, we generated data for one day at a sampling rate of 0.5\,Hz.
The LISA response to the GW for the six links between the three SC was then loaded by the MATLAB script introduced in~\sref{sec:TTL_simulation}. 

\paragraph{Stochastic gravitational wave background from stellar-origin binary black holes} 
The superposition of several sources such as relic GWs in the earliest moments of the universe can generate the SGWB. Binary black hole and binary neutron star mergers throughout the history of the universe can also create SGWB~\cite{Redbook2024,Christensen_2019}. We chose SOBBHs to simulate SGWB~\cite{Babak_2023}. Its energy density is modeled as
\begin{equation}
    h^{2}\Omega_{\rm GW}=h^{2}\Omega_{\rm GW}(f_{0})\left(\frac{f}{f_{0}} \right)^{2/3}
\end{equation}
where we assumed $\Omega_{\rm GW}(f_{0})=1.15\times10^{-12}$ and $f_{0}$=3\,mHz, which is the upper limit given in~\cite{Babak_2023}.
Its spectrum is shown in figure~\ref{fig:ASD_gw} in light blue, which corresponds to the data generated with the LISA GW response package~\cite{lisagw}. It is below the instrument noise and, consequently, should not have any noticeable effect on the TTL recovery algorithms.

\paragraph{Resolvable galactic binaries} 
These are the most numerous sources in the LISA band: about $10^{7}$ are expected to radiate with periods from hours to minutes. Their strain amplitudes are rather weak compared to black hole binaries signals. Nevertheless, about $10^{4}$ monochromatic individual sources will be detectable after long integration times~\cite{amaro2023}. We use the GB model described in~\cite{lisagw}. Their parameters are taken from LISA Cat tools~\cite{LISAcat}, which include amplitude, frequency, initial phase of the observer ($\phi$), polarization ($\psi$), inclination ($i$), and ecliptic latitude and longitude ($\beta$, $\lambda$).
We chose the loudest 100 signals and simulate them for one day. The strain levels are around 10$^{-22}$ and the frequencies span from 1\,mHz to 20\,mHz. The ASDs of these signals are shown in green in figure~\ref{fig:ASD_gw}, where one can observe some signals slightly above the instrument noise.

\paragraph{Galactic Confusion Noise}
A large blend of unresolvable GW sources make up a noise source known as double white dwarf (WDWD) galactic confusion noise or simply galactic foreground~\cite{Ruiter_2010,Robson_2017}.
This foreground limits LISA's sensitivity to other types of resolvable sources in the milli-hertz band and it is usually considered as a noise source since the sources cannot be resolved individually. 
We generated the confusion noise by simulating 500 WDWD-like galactic binaries by referencing parameters from LISAcat~\cite{LISAcat}. The amplitude and frequency parameters are drawn from truncated Gaussian distributions where the strain spans from $10^{-24}$ to $10^{-22}$ with a mean value as well as standard deviation of $10^{-23}$. 
The frequencies range from 0.1\,mHz to 1\,Hz with a mean value of 1\,mHz and a standard deviation of 1\,mHz.
As for the orientation parameters: ecliptic latitude ($\beta$), ecliptic longitude ($\lambda$), inclination ($i$), observer phase ($\phi$), and polarization angle ($\psi$), we chose random values between 0 and $\pi$. Finally, the strain (in Doppler shifts) of the 500 sources was re-scaled in order to achieve the expected foreground spectral signature as $h_{i,\rm scaled} = ({0.02}/{f_{i}})h_{i}$. 
The resulting confusion noise is shown in \fref{fig:ASD_gw} (yellow trace), which, despite our simplified modeling method, is in good agreement with the expected one~\cite{Redbook2024}, and accurate enough for our purposes. \Fref{fig:time_gw} shows the confusion noise in the time-domain compared to the sum of all considered GW signals.

\paragraph{Extreme mass ratio inspirals}
These are astrophysical phenomena where a much smaller compact object eventually plunges into a much larger object (such as a massive black hole)~\cite{babak2017}. We generated them using LDC tools~\cite{ldc} and chose a MBH of $10^{6}M_{\odot}$ and a companion of 23$M_{\odot}$ at a distance of 3\,Gpc. The response is shown in orange in figure~\ref{fig:ASD_gw}, and its amplitude is well below the LISA instrument noise for the measurement time of one day and should therefore have no effect on the recovery of the TTL coefficients.

\paragraph{Massive Black Hole Binaries} 
We generated signals for four different MBHBs mergers. As previously, we only considered one day of observation. Unlike the previous sources, MBHBs can exhibit high signal-to-noise ratios even for short observation times. We simulated the MBHB signals using LDC tools~\cite{ldc}, where we used the masses, frequency, and distances given in table~\ref{tab:gw_signals}. These strain waveforms were then passed to the LISA GW response package~\cite{lisagw}. MBHB\,1 and 4 merged in the middle of the day, at the 12 hour mark, while MBHB\,2 merged at the 18 hour mark. MBHB\,3 merged at the end of the 24-hour period. The ASDs of the signals are shown in figure~\ref{fig:ASD_gw} in purple, where the mergers are clearly above the instrumentation noise and cover a wide frequency range: from 0.6\,mHz to 0.1\,Hz. The dip at about 75\,mHz in \fref{fig:ASD_gw} (top) is a consequence of the LISA detector's transfer function~\cite{lobo}. \Fref{fig:time_gw} shows the mergers in time domain.

\subsection{Uncertainty of the estimators}

\begin{figure}
\begin{indented}
\item[]\includegraphics[width=0.9\textwidth]{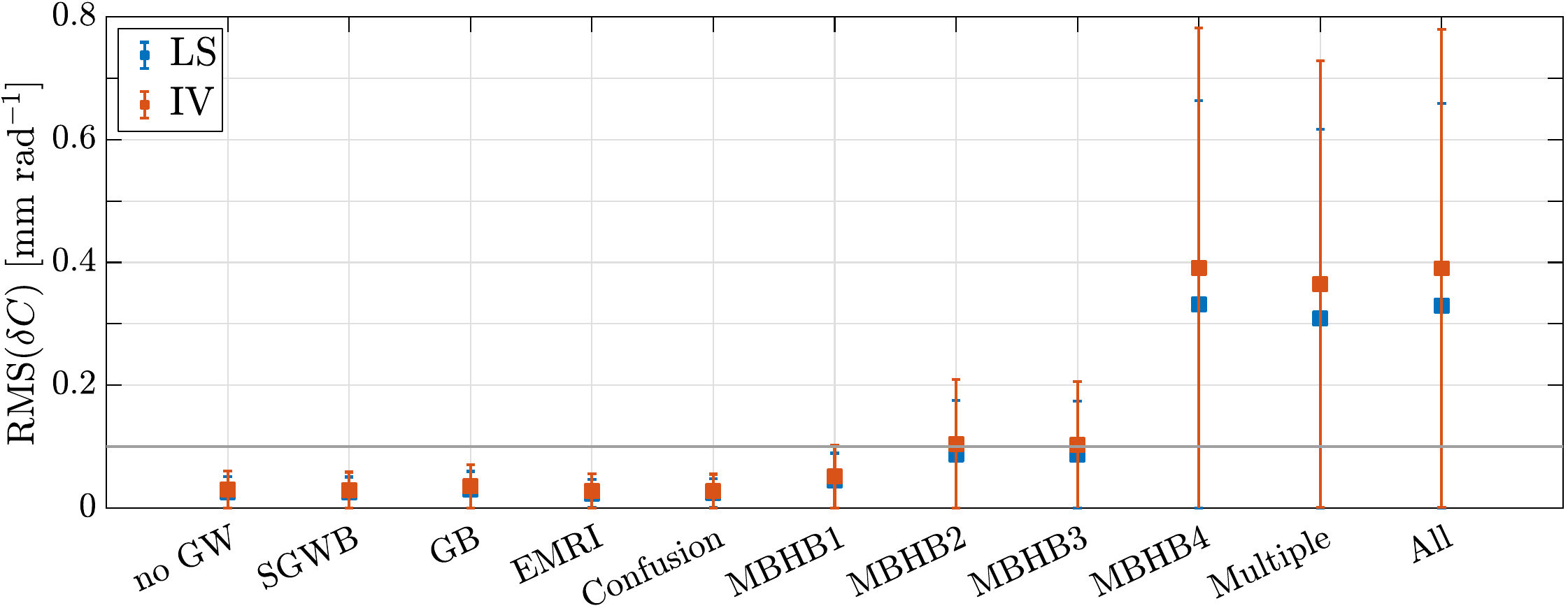}
  \caption{Root mean square (RMS) of the coefficient estimation error $\delta C$ in the presence of different GW signals. Each RMS value was computed considering 100 data sets with different noise and jitter realizations, but the same GW signal. 
  In the second last case labeled `Multiple', we considered all GW signals, but only the largest MBHB (`MBHB\,4').
  In the last case labeled `All', all introduced GW signals were included.
  The error bars are the computed standard deviations. The gray line highlights the 0.1\,mm/rad deviation level.}
  \label{fig:RMS_gwsignals}
\end{indented}
\end{figure}

The results of the coefficient estimations considering the GW signals are shown in \fref{fig:RMS_gwsignals}.
For the nominal DWS jitter levels, the LS estimator performs slightly better than the IV method. However, the results of the two estimators are close in all cases. 
We see differences between the different GW signals scenarios. 
The SOBBH SGWB, resolvable GB, and EMRI amplitudes are small for one day of data. Therefore, their effect on the estimator accuracy is negligible. 
The unresolved galactic binaries (confusion noise) are dominant only at low frequencies and likewise have little effect on the coefficient estimates.
We find larger estimation errors and variations in cases of the MBHB mergers and if considering multiple GW sources.
These GW signals add significant content to the interferometric length measurements at frequencies where TTL coupling is also dominant, see~\fref{fig:ASD_gw}.
Particularly the last merger type, MBHB\,4, leads to a significantly larger coefficient estimation error. It exceeds the 0.1\,mm/rad error requirement by more than a factor of three. We see in \fref{fig:ASD_gw} that this merger has the largest response in the single link and its frequency band overlaps with the frequency range considered for the TTL coupling coefficient estimation.
When adding all GW signals (case labeled `All' in \fref{fig:RMS_gwsignals}), we find an estimation error of the same magnitude.
We get a similar result if we include all signal types, but only the largest MBHB merger (case labeled `Multiple' in the same figure).
All other LS estimation errors are below 0.1\,mm/rad for nominal instrument noise and jitter levels.

\section{Summary}
\label{sec:summary}

In the present work, we have investigated the dependence of the coupling coefficient estimation error on the angular jitter and readout noise characteristics and on exemplary GW signal sources. 
An accurate coefficient estimation is crucial for the TTL noise mitigation strategy via realignment.
Furthermore, the estimation errors also add to the TTL noise residual after subtraction.
The main results of our performance analysis are presented in \sref{sec:TTL_uncertainty} and \sref{sec:TTL_GW}.

We have shown (\sref{sec:TTL_uncertainty}) that the LS coefficient estimate becomes significantly biased for increasing levels of readout noise. The bias depends on the coupling coefficients, the jitter characteristics, and the DWS noise level. 
It can be predicted analytically (\sref{sec:TTL_estimators}). This bias equation also provides a handle to investigate the effect of a chosen or observed jitter shape or DWS readout noise level on the coefficient error without running long simulations.
On the downside, the analytical model is very sensitive to its input parameters. Changing the DWS readout noise levels by 10\,\% will change the bias by about the same percentage. Therefore, these levels must be known fairly well to be useful for in-flight correction of the LS bias estimate. 

An alternative coefficient computation method is the IV estimator.
This estimator is not systematically biased by the DWS readout noise. However, this readout noise can have an effect on the variance of the estimation results depending on the number of data points. Also, the IV method does not converge if the jitter is white, which we assume to be unrealistic in LISA. 
Therefore, this estimator holds as a good alternative to the LS estimator for the TTL coupling coefficient computation in the case of large DWS readout noise levels.
For nominal noise and jitter levels, the LS estimator is the preferred method since its error and variance are smaller given 1\,day (2000\,s for the maneuvers' case) of data and a sampling frequency of 0.5\,Hz (4\,Hz for the maneuvers' case). 

We also analyzed the performance of the estimators for different levels of SC jitter. In this case, the LS and IV estimators show a bias at low levels of SC jitter. 
As the SC jitter increases, the error of the pitch coefficient decreases. However, the yaw jitter on the same SC becomes highly correlated, which limits the accuracy of the corresponding coupling coefficients. Therefore, increasing the SC jitter only marginally reduces the RMS of the coefficient estimation errors. 
This correlation is not described by the analytical bias equation. 

Finally, in \sref{sec:TTL_GW} we examined the coefficient estimation error in the presence of several GWs. Most of these are too small for one day of data, or are only significant at frequencies below the frequency range considered for TTL coupling coefficient estimation.
However, we find a significant effect in the case of MBHB mergers. 
The most violent one, with $10^6$ solar masses each at redshift $z$=2, gives a coefficient estimation error of 0.33\,mm/rad.
We conclude that in the presence of large mergers, the data cannot be used to estimate the TTL coupling coefficient. We would have to cut out the merger from the time series data considered in the TTL noise fit, or adjust the frequency range used in the fit to exclude the merger signal. 
In both cases we lose data, which reduces the accuracy and certainty of the LS and IV estimators.

Overall, we have presented here a comprehensive study of the dependencies of TTL coefficient estimation errors and biases on DWS readout level, SC jitter levels and shapes, and a variety of GW signals. 
We have identified the scenarios in which the TTL coefficients can be estimated well and those in which they become less reliable. 
While we do not yet know what performance LISA will achieve in space, our analysis can serve as a reference for TTL coupling coefficient estimation strategies.

\ack{
\label{sec:acc}
	We thank P.~Fulda for his support and the valuable discussions on the content of this paper.
	We thank H.~Wegener for helpful conversations on the TTL coupling maneuver design.
	We also wish to thank the MAE/Physics LISA simulation group at the University of Florida for lively discussions and valuable feedback.
    Different simulators have been used in this work for data generation and cross-checks.
    We thank M.~Hewitson for providing the MATLAB simulator LISASim.
    We acknowledge J-B.~Bayle, O.~Hartwig, A.~Petiteau and M.~Lilley for developing and maintaining LISANode and thank O.~Sauter for his enduring support.
    We acknowledge the LDC working group for providing the LISA Data Challenge software.
	For the LISA GW Response package, we acknowledge its authors J-B.~Bayle, Q.~Baghi, A.~Renzini and M.~Le Jeune, as well as the LISA Simulation Expert Group.
    M.~Hartig acknowledges the support by the PRIME programme of the German Academic Exchange Service (DAAD) with funds from the German Federal Ministry of Education and Research (BMBF).
    S.~Paczkowski gratefully acknowledges support by the Deutsches Zentrum für Luft- und Raumfahrt (DLR, German Space Agency) with funding of the Federal Ministry for Economic Affairs and Climate Action (BMWK) with a decision of the Deutsche Bundestag (DLR Project Reference No.~FKZ 50OQ2301 based on funding from FKZ 50OQ1801).
}

\bibliographystyle{unsrt}
\section*{References}
\bibliography{References.bib}

\appendix

\section{Interferometer noises used in our analysis}
\label{app:ifo_noise}

For our simulations, we implemented the ASDs of the following interferometer length measurement noises. The amplitudes and frequency shapes were taken from recent publications~\cite{Paczkowski2022,George2022} or correspond to the implementations in other LISA simulators \cite{Bayle2023,LISANode,LISASim}. 
\begin{itemize}
    \item Fiber (or back-link) noise:
\begin{eqnarray}
  S^{\mathrm{FIB},1/2} 
  = 3\,\mathrm{pm}/\sqrt{\mathrm{Hz}}\
  \sqrt{ 1+ \left(\frac{2\,\mathrm{mHz}}{f}\right)^4 }
\end{eqnarray}

\item Test mass acceleration noise (converted to displacement noise):
\begin{eqnarray}
  S^{\mathrm{TMF},1/2} 
  = \frac{2.4}{(2\pi\,f)^2}\,\mathrm{fm}/\mathrm{s}^2/\sqrt{\mathrm{Hz}}\
  \sqrt{1+ \left(\frac{0.4\,\mathrm{mHz}}{f}\right)^2}\ 
  \sqrt{1+ \left(\frac{8\,\mathrm{mHz}}{f}\right)^4}
\end{eqnarray}

\item Inter-satellite interferometer readout noise:
\begin{eqnarray}
  S^{\mathrm{ISR},1/2} 
  = 6.35\,\mathrm{pm}/\sqrt{\mathrm{Hz}}\
  \sqrt{ 1+ \left(\frac{2\,\mathrm{mHz}}{f}\right)^4 }
\end{eqnarray}

\item Test mass interferometer readout noise:
\begin{eqnarray}
  S^{\mathrm{TMR},1/2} 
  = 1.42\,\mathrm{pm}/\sqrt{\mathrm{Hz}}\
  \sqrt{ 1+ \left(\frac{2\,\mathrm{mHz}}{f}\right)^4 }
\end{eqnarray}

\item Reference interferometer readout noise:
\begin{eqnarray}
  S^{\mathrm{RFR},1/2} 
  = 3.32\,\mathrm{pm}/\sqrt{\mathrm{Hz}}\
  \sqrt{ 1+ \left(\frac{2\,\mathrm{mHz}}{f}\right)^4 }
\end{eqnarray}
\end{itemize}

We did not include the laser frequency noise in our analysis. While being the largest noise source in the long arm measurements, its residual after TDI is significantly smaller than the other considered noises.
Note that the noise in the local and the far SC's test mass interferometers add into the single link measurement in LISA. 
See \cite{Nam2023} for more details on the noise PSDs after TDI.

\section{Bias of the least squares estimator for colored jitter}
\label{app:bias}

Here we present the equations for the bias of the LS estimator. While the equations~\eref{eq:dC_etaRX}-\eref{eq:dC_phiTX} hold for the case of white jitter, we extend them here for the more general case of arbitrary jitter shapes. 
Still, we consider the jitter of the three SC and the six MOSAs to be uncorrelated.
\begin{eqnarray}\hspace*{-5ex}
  \lim_{N\rightarrow\infty} \delta \widehat{C}^\mathrm{LS}_{ij \eta \rm RX} 
  &=& K_{\eta1} \cdot\left( C_{ij \eta \rm RX} +C_{ik \eta \rm RX} -C_{ij \eta \rm TX} -C_{ik \eta \rm TX} \right) \nonumber\\
  &+& K_{\eta2} \cdot\left( C_{ij \eta \rm RX} -C_{ik \eta \rm RX} -C_{ij \eta \rm TX} +C_{ik \eta \rm TX} \right) \nonumber\\
  &+& K_{\eta3} \cdot\left( C_{ij \eta \rm RX} +C_{ik \eta \rm RX} +C_{ij \eta \rm TX} +C_{ik \eta \rm TX} \right) \nonumber\\
  &+& K_{\eta4} \cdot\left( C_{ij \eta \rm RX} -C_{ik \eta \rm RX} +C_{ij \eta \rm TX} -C_{ik \eta \rm TX} \right) \,,
  \label{eq:dC_etaRX_app}  
  \\ \hspace*{-5ex}
  \lim_{N\rightarrow\infty} \delta \widehat{C}^\mathrm{LS}_{ij \eta \rm TX} 
  &=& K_{\eta1} \cdot\left( -C_{ij \eta \rm RX} -C_{ik \eta \rm RX} +C_{ij \eta \rm TX} +C_{ik \eta \rm TX}  \right) \nonumber\\
  &+& K_{\eta2} \cdot\left( -C_{ij \eta \rm RX} +C_{ik \eta \rm RX} +C_{ij \eta \rm TX} -C_{ik \eta \rm TX}  \right) \nonumber\\
  &+& K_{\eta3} \cdot\left( C_{ij \eta \rm RX} +C_{ik \eta \rm RX} +C_{ij \eta \rm TX} +C_{ik \eta \rm TX}  \right) \nonumber\\
  &+& K_{\eta4} \cdot\left( C_{ij \eta \rm RX} -C_{ik \eta \rm RX} +C_{ij \eta \rm TX} -C_{ik \eta \rm TX}  \right) \,,
  \label{eq:dC_etaTX_app} 
  \\ \hspace*{-5ex}
  \lim_{N\rightarrow\infty} \delta \widehat{C}^\mathrm{LS}_{ij \varphi \rm RX} 
  &=& K_{\varphi1} \cdot\left( C_{ij \varphi \rm RX} +C_{ik \varphi \rm RX} -C_{ij \varphi \rm TX} -C_{ik \varphi \rm TX}  \right) \nonumber\\
  &+& K_{\varphi2} \cdot\left( C_{ij \varphi \rm RX} -C_{ik \varphi \rm RX} -C_{ij \varphi \rm TX} +C_{ik \varphi \rm TX}  \right) \nonumber\\
  &+& K_{\varphi3} \cdot\left( C_{ij \varphi \rm RX} +C_{ik \varphi \rm RX} +C_{ij \varphi \rm TX} +C_{ik \varphi \rm TX}  \right) \nonumber\\
  &+& K_{\varphi4} \cdot\left( C_{ij \varphi \rm RX} -C_{ik \varphi \rm RX} +C_{ij \varphi \rm TX} -C_{ik \varphi \rm TX}  \right) \,,
  \label{eq:dC_phiRX_app} 
  \\ \hspace*{-5ex}
  \lim_{N\rightarrow\infty} \delta \widehat{C}^\mathrm{LS}_{ij \varphi \rm TX} 
  &=& K_{\varphi1} \cdot\left( -C_{ij \varphi \rm RX} -C_{ik \varphi \rm RX} +C_{ij \varphi \rm TX} +C_{ik \varphi \rm TX}  \right) \nonumber\\
  &+& K_{\varphi2} \cdot\left( -C_{ij \varphi \rm RX} +C_{ik \varphi \rm RX} +C_{ij \varphi \rm TX} -C_{ik \varphi \rm TX}  \right) \nonumber\\
  &+& K_{\varphi3} \cdot\left( C_{ij \varphi \rm RX} +C_{ik \varphi \rm RX} +C_{ij \varphi \rm TX} +C_{ik \varphi \rm TX}  \right) \nonumber\\
  &+& K_{\varphi4} \cdot\left( C_{ij \varphi \rm RX} -C_{ik \varphi \rm RX} +C_{ij \varphi \rm TX} -C_{ik \varphi \rm TX}  \right) \,,
  \label{eq:dC_phiTX_app} 
\end{eqnarray}
with
\begin{description}
  \item[$K_{\eta1}$]= $\frac{3\,\sdws/2}{(6-8\rmeII-8\rmeIV+8\rmeVI-2\rmeVIII)\,\sme+(3-4\rseII-4\rseIV+4\rseVI-\rseVIII)\,\sse+6\,\sdws}$
  \item[$K_{\eta2}$]= $\frac{3\,\sdws/2}{(6-8\rmeII-8\rmeIV+8\rmeVI-2\rmeVIII)\,\sme+(9-12\rseII-12\rseIV+12\rseVI-3\rseVIII)\,\sse+6\,\sdws}$
  \item[$K_{\eta3}$]= $\frac{13\,\sdws/2}{(26-8\rmeII-24\rmeIV+8\rmeVI+2\rmeVIII)\,\sme+(25-8\rseII-20\rseIV+8\rseVI+\rseVIII)\,\sse+26\,\sdws}$
  \item[$K_{\eta4}$]= $\frac{13\,\sdws/2}{(26-8\rmeII-24\rmeIV+8\rmeVI+2\rmeVIII)\,\sme+(27-8\rseII-28\rseIV+8\rseVI+3\rseVIII)\,\sse+26\,\sdws}$
  \item[$K_{\varphi1}$]= $\frac{3\,\sdws/4}{(3-4\rmpII-4\rmpIV+4\rmpVI-\rmpVIII)\,\sme+3\,\sdws}$
  \item[$K_{\varphi2}$]= $\frac{3\,\sdws/4}{(3-4\rmpII-4\rmpIV+4\rmpVI-\rmpVIII)\,\sme+(6-8\rspII-8\rspIV+8\rspVI-2\rspVIII)\,\sse+3\,\sdws}$
  \item[$K_{\varphi3}$]= $\frac{13\,\sdws/4}{(13-4\rmpII-12\rmpIV+4\rmpVI+\rmpVIII)\,\sme+(12-4\rspII-8\rspIV+4\rspVI)\,\sse+13\,\sdws}$
  \item[$K_{\varphi4}$]= $\frac{13\,\sdws/4}{(13-4\rmpII-12\rmpIV+4\rmpVI+\rmpVIII)\,\sme+(14-4\rspII-16\rspIV+4\rspVI+2\rspVIII)\,\sse+13\,\sdws}$
\end{description}
and 
\begin{description}
  \item[$\rho^{k\tau,\mathrm{B}}_\alpha$] the normalized auto-correlation coefficient of the jitter $\alpha\in\{\varphi,\eta\}$ of B\,$\in\{\mathrm{SC},\mathrm{MO}\}$ and its by $k\in\{2,4,6,8\}$ times the inter-satellite propagation time delayed series.
\end{description}

The normalized auto-correlations $\rho^{k\tau,\mathrm{B}}_\alpha$ and the variances $\sigma^\mathrm{2B}_\alpha$ can be computed from the data time series or, more generally, from the jitter and readout noise PSDs. 
In our work, we computed these values from the PSD.
The variance is associated to the jitter PSD $S^\mathrm{B}_\alpha$ by Parseval's theorem
\begin{eqnarray}
  \sigma^\mathrm{2B}_\alpha = \int_{-\infty}^\infty S^\mathrm{B}_\alpha\ \mathrm{d}f \,,
\end{eqnarray}
and the normalized auto-correlation can be computed using the Wiener-Khinchin theorem
\begin{eqnarray}
  \rho^{k\tau,\mathrm{B}}_\alpha
  = \int_{-\infty}^\infty S^\mathrm{B}_\alpha\ \cos(2\pi\,f\cdot k\,\overline{\tau})\ \mathrm{d}f \left/ 
  \int_{-\infty}^\infty S^\mathrm{B}_\alpha\ \mathrm{d}f \right. \,,
\end{eqnarray}
with $\overline{\tau}$ being the average time delay along one LISA arm.

\section{Correlations between the coefficient estimates}
\label{app:correlation}

\begin{figure}
\begin{indented}
\item[]\includegraphics[width=0.45\textwidth]{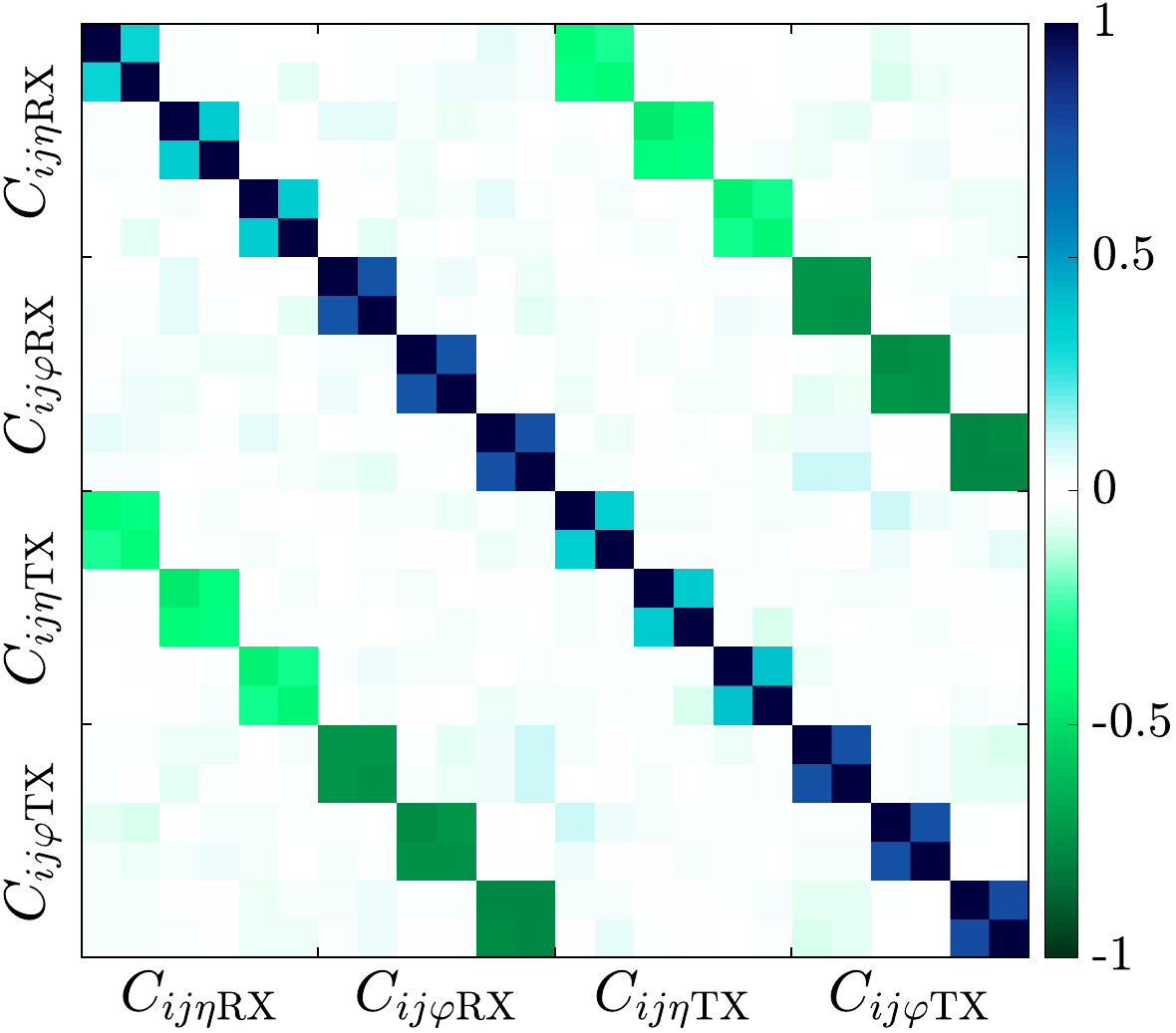}\hfill
  \includegraphics[width=0.45\textwidth]{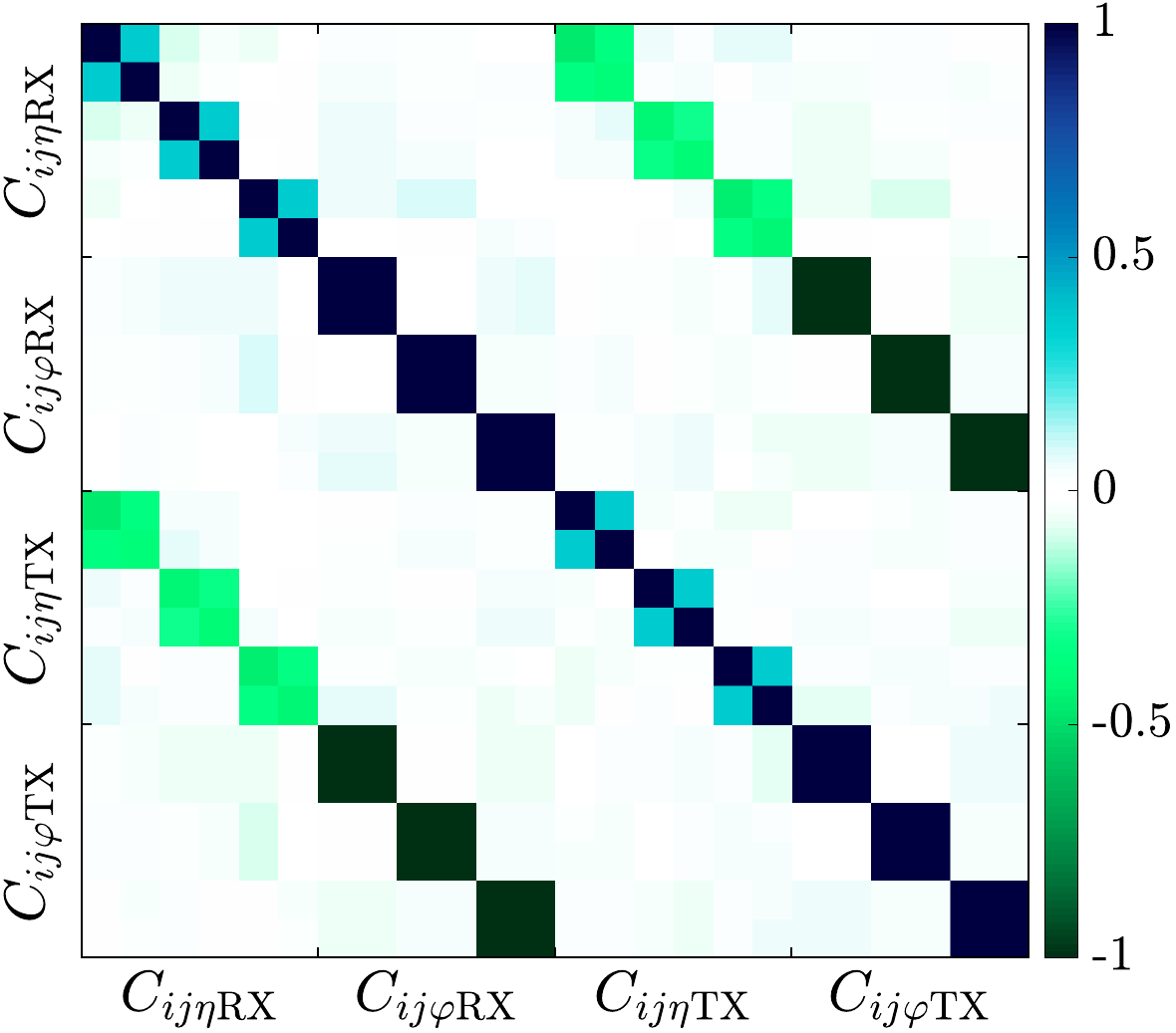}
  \caption{Correlation matrix for the 24 LS coupling coefficient estimates. Values close to one denote a high correlation. Values close to -1 indicate a high anti-correlation. The matrix was computed using the coefficient estimates for 1000 correlations.
  Left: Correlations for nominal jitter and noise settings.
  Right: Correlations for 10 times increased SC jitter.
  The order of the coefficient indices is $ij\in\{12,13,23,21,31,32\}$.}
  \label{fig:correlation}
\end{indented}
\end{figure}

Some of the coefficient estimates are correlated. 
We show this in \fref{fig:correlation} for the LS estimator.
We see that the coefficients for pitch and yaw are correlated on the same SC. The same is true for the RX coefficients and their TX counterparts. 
Comparing the left (nominal settings) and right (10 times increased SC jitter) plots in \fref{fig:correlation}, we see that the correlations of the yaw coefficients have increased significantly due to the increase in SC jitter. 

\section{Estimator performance for random TTL coefficients}
\label{app:rand_TTL}

\begin{figure}
\begin{indented}
\item[]\includegraphics[width=0.45\textwidth]{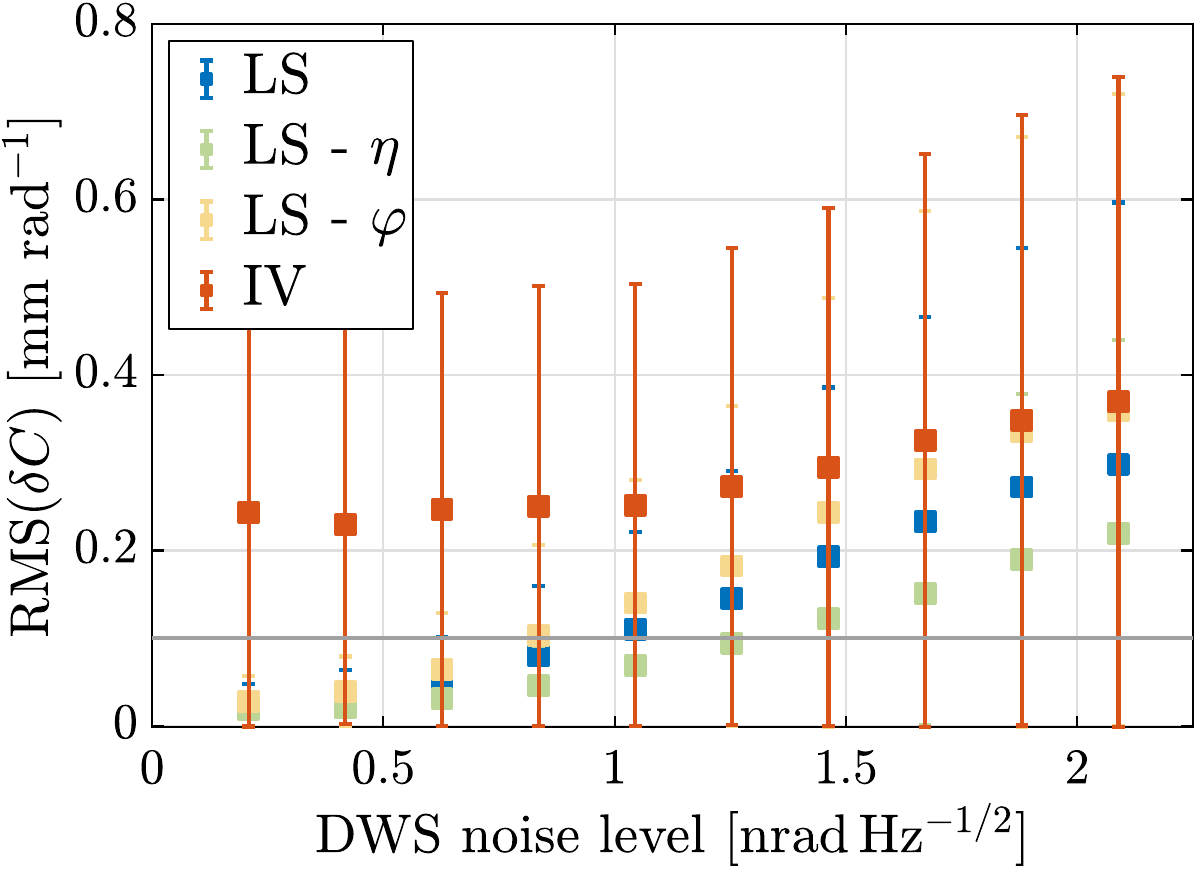}\hfill
  \includegraphics[width=0.45\textwidth]{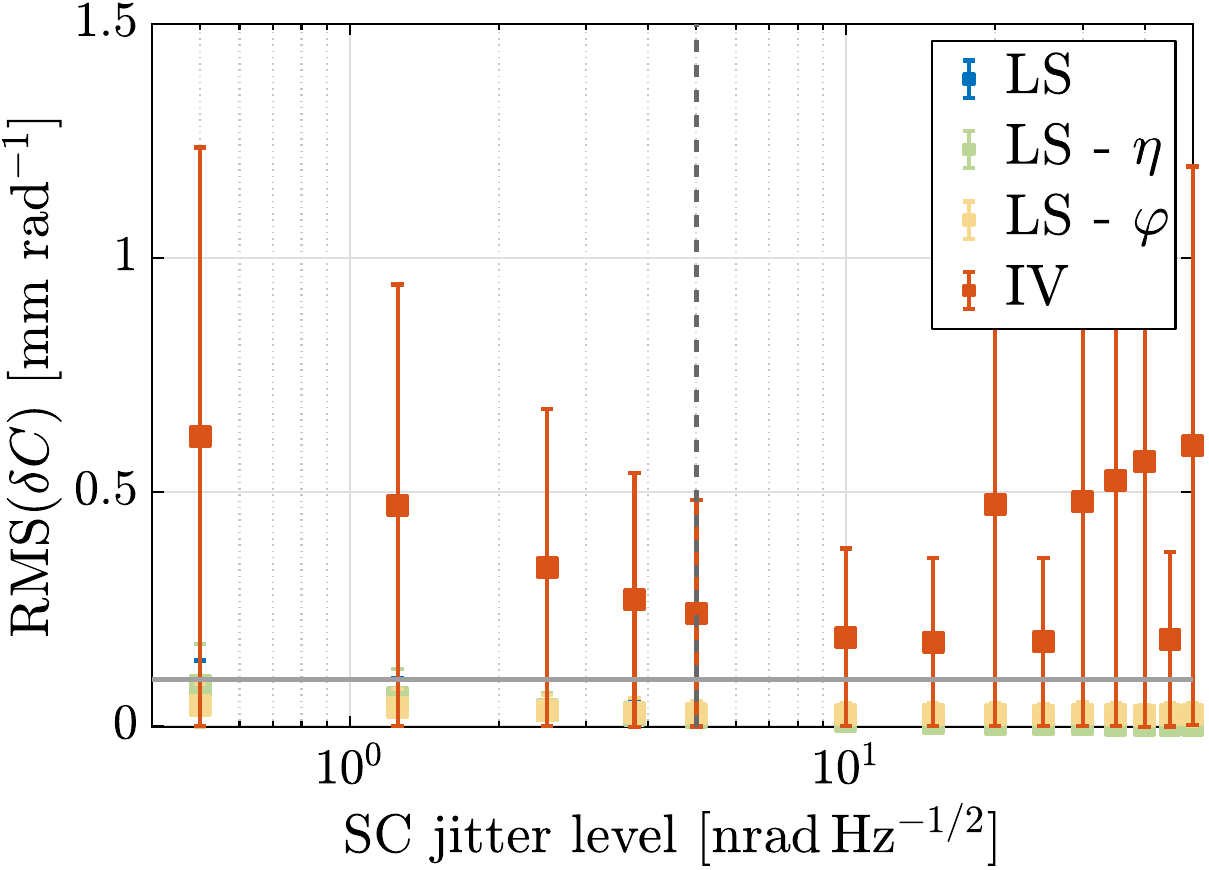}
  \caption{RMS of the coefficient estimation error $\delta C$ for arbitrary TTL coupling coefficients in the case of white jitter. 
  Left: RMS of the estimation errors for increased DWS readout noise levels.
  Right: RMS of the estimation errors for different SC jitter levels.
  Each RMS value was computed considering 100 data sets with different noise and jitter realizations. The error bars are the computed standard deviations.
  The gray dashed vertical line marks the nominal SC jitter level. The gray horizontal line highlights the 0.1\,mm/rad deviation level.}
  \label{fig:RMS_white_rand}
\end{indented}
\end{figure}

\begin{figure}
\begin{indented}
\item[]\includegraphics[width=0.45\textwidth]{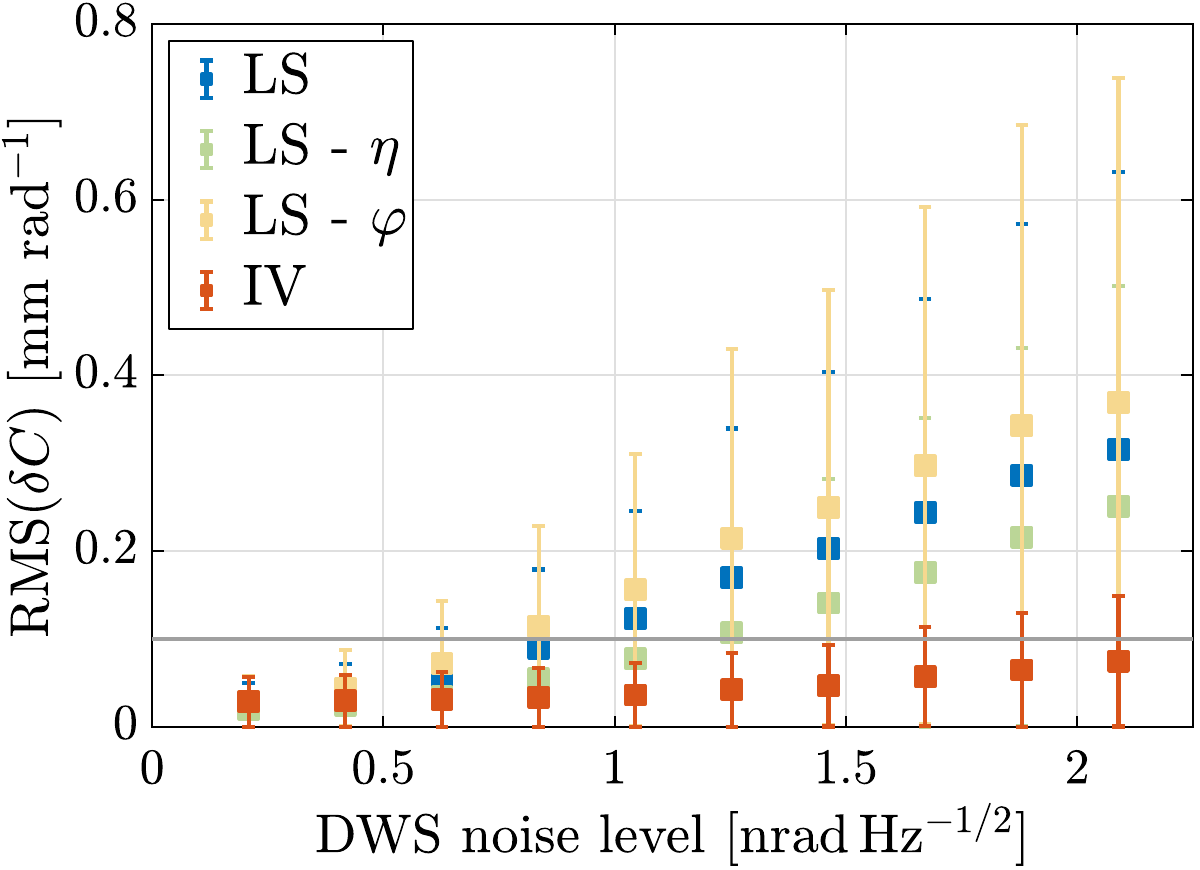}\hfill
  \includegraphics[width=0.45\textwidth]{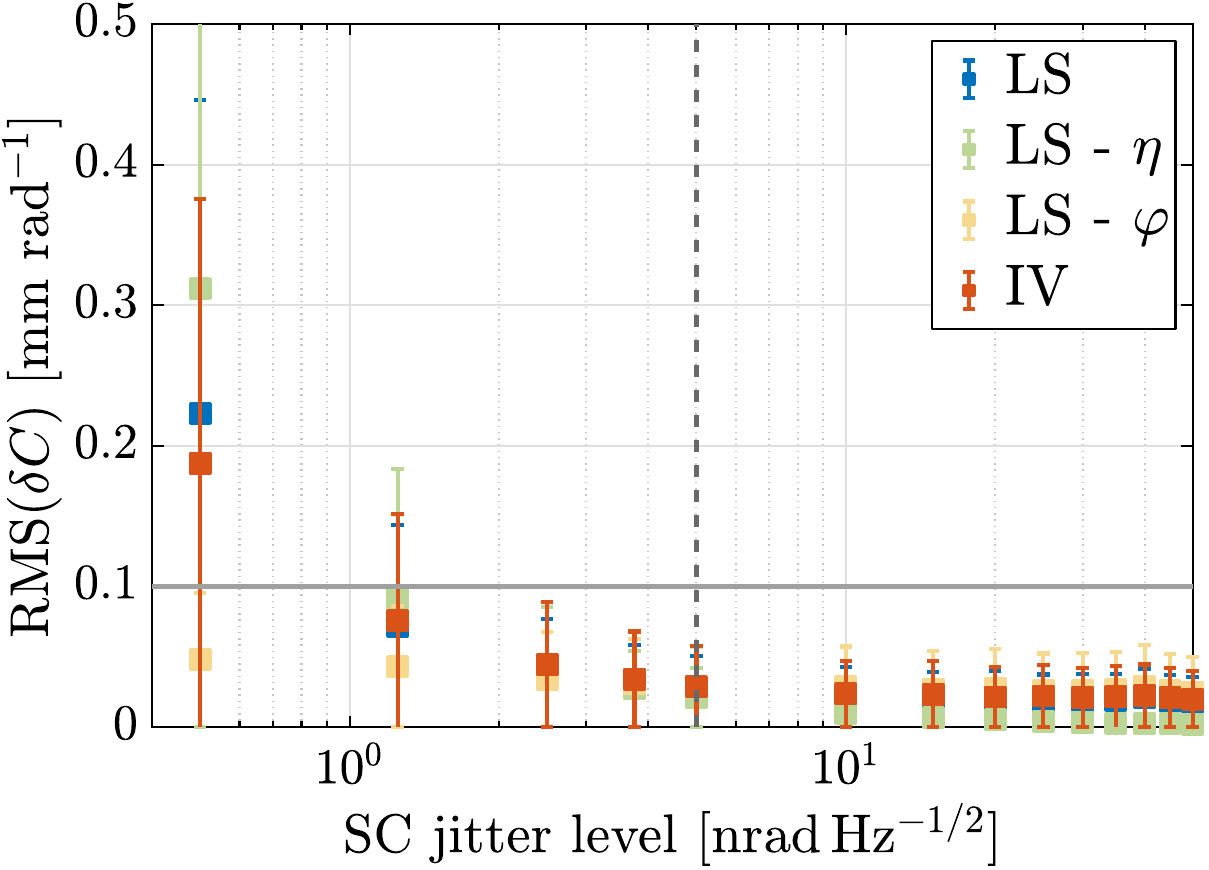}
  \caption{RMS of the coefficient estimation error $\delta C$ for arbitrary TTL coupling coefficients in the case of colored jitter. 
  Left: RMS of the estimation errors for increased DWS readout noise levels.
  Right: RMS of the estimation errors for different SC jitter levels.
  Each RMS value was computed considering 100 data sets with different noise and jitter realizations. The error bars are the computed standard deviations.
  The gray dashed vertical line marks the nominal SC jitter level. The gray horizontal line highlights the 0.1\,mm/rad deviation level.}
  \label{fig:RMS_dynamic_rand}
\end{indented}
\end{figure}

\begin{figure}
\begin{indented}
\item[]\includegraphics[width=0.45\textwidth]{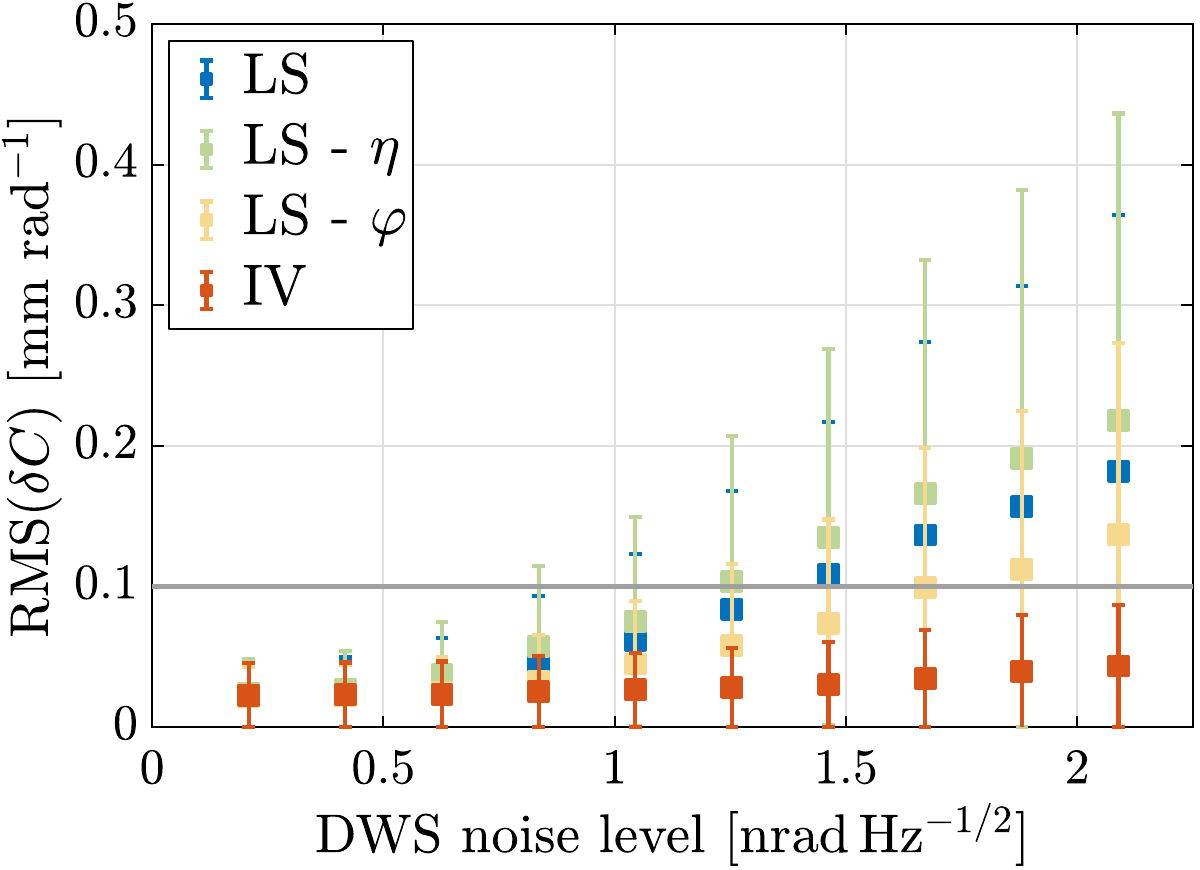}
  \caption{RMS of the coefficient estimation error $\delta C$ for arbitrary TTL coupling coefficients and increased DWS readout noise levels in the case of maneuvers. For the LS estimator, we show the error of the pitch ($\eta$) and yaw ($\varphi$) coefficients separately.
  The ten data points correspond to DWS noise levels increased by factors of one to ten. Each RMS value was computed considering 100 data sets with different noise and jitter realizations. The error bars are the computed standard deviations. The gray line highlights the 0.1\,mm/rad deviation level.}
  \label{fig:RMS_maneuver_rand}
\end{indented}
\end{figure}

Here we show the performance of the LS and IV estimators for arbitrary TTL coupling coefficients with $\left\vert C^\mathrm{TTL}_i\right\vert\leq2.3\,$mm/rad (figures \ref{fig:RMS_white_rand}, \ref{fig:RMS_dynamic_rand}, \ref{fig:RMS_maneuver_rand}).
We use the same settings as in the above investigations. We run 100 simulations with different jitter and noise realizations for each analyzed case. In addition, we choose an arbitrary set of TTL coupling coefficients in each simulation. 
In each of the figures, we plotted the RMS of the LS and IV estimator errors. 
Additionally, we looked at the RMS for the LS pitch ($\eta$) and yaw ($\varphi$) coefficient estimates. Note that as the DWS readout noise increases, the yaw coefficient errors are larger for white and colored jitter. This is consistent with the prediction of the analytically computed bias. 
However, when we apply the maneuvers, the yaw coefficients are more accurate and precise than the pitch coefficients. 
This is due to the fact that the yaw MOSA maneuvers are significantly larger than before. In addition to improving the coefficient estimation, these maneuvers also reduce the correlations between the yaw jitters on the same SC.

\end{document}